%% file: main.tex
\documentclass[12pt]{article}
\usepackage{graphicx}
\usepackage{natbib} 
\usepackage{url} 

\newcommand{\blind}{0}

\newcommand{\st}{\mbox{such that}}
\usepackage{comment}

\def\hb#1{#1}

\include{macros}
\usepackage{diagbox}
\usepackage{bbm}
\usepackage{multirow}
\usepackage[english]{babel}
\usepackage{amsthm}
\newtheorem{theorem}{Theorem}[section]

\newtheorem{lemma}[theorem]{Lemma}

\usepackage[colorlinks=true,    
            linkcolor=blue,     
            citecolor=red,      
            urlcolor=magenta]{hyperref}

\makeatletter

\begin{document}

\def\spacingset#1{\renewcommand{\baselinestretch}%
{#1}\small\normalsize} \spacingset{1}


\if0\blind
{
  \title{\bf Batch Sequential Experimental Design for Calibration of Stochastic Simulation Models}
  \author{\"{O}zge S\"{u}rer\thanks{
    The author gratefully acknowledges OAC 2004601}\hspace{.2cm}\\
    Department of Information Systems \& Analytics, Miami University\\}
  \maketitle
} \fi

\if1\blind
{
  \bigskip
  \bigskip
  \bigskip
  \begin{center}
    {\LARGE\bf Batch Sequential Experimental Design for Calibration of Stochastic Simulation Models}
\end{center}
  \medskip
} \fi

\bigskip
\begin{abstract}
    Calibration of expensive simulation models involves an emulator based on simulation outputs generated across various parameter settings to replace the actual model.
    Noisy outputs of stochastic simulation models require many simulation evaluations to understand the complex input-output relationship effectively.
    Sequential design with an intelligent data collection strategy can improve the efficiency of the calibration process.
    The growth of parallel computing environments can further enhance calibration efficiency by enabling simultaneous evaluation of the simulation model at a batch of parameters within a sequential design.
    This article proposes novel criteria that determine if a new batch of simulation evaluations should be assigned to existing parameter locations or unexplored ones to minimize the uncertainty of posterior prediction.
    Analysis of several simulated models and real-data experiments from epidemiology demonstrates that the proposed approach results in improved posterior predictions.
\end{abstract}

\noindent%
\vfill

\newpage
\spacingset{1} 

\section{Introduction}

Simulation models are essential tools to analyze or predict the behavior of complex systems in engineering and science disciplines. A simulation model takes an input and generates the corresponding output. Unlike deterministic simulation models, stochastic simulation models generate different outputs when evaluated repeatedly at the same input due to the randomness present in the model. Examples include the simulation of nuclear dynamics \citep{Liyanage2023}, an Ebola epidemic \citep{Fadikar2018}, and a biological model of mitochondrial DNA population dynamics \citep{Henderson2009}. Simulation models often involve unknown input parameters that govern their behavior. Calibration is a way to estimate these unknown parameters using observed real-world data \citep{Sung2024}. Calibration becomes a more challenging problem if evaluating a simulation model with an input is computationally expensive. This problem is compounded in stochastic simulation models since noisy simulation outputs require many simulation runs to distinguish signal from noise.

An emulator is often built as a cost-effective surrogate for computationally demanding simulation models. The Gaussian process (GP) model is a widely employed surrogate model for deterministic simulation models \citep{Rasmussen2005, gramacy2020surrogates}. GPs provide not only precise predictions of simulation models but also the measure of uncertainty. GPs are constructed using a simulation data set involving inputs (referred to as design) and the associated simulation outputs. Typical methods for creating designs include uniform sampling and space-filling designs like Latin hypercube sampling (LHS) \citep{McKay1979}; refer to \cite{santner2018design} for an extensive overview. Although GP models can be extended for modeling stochastic simulation outputs, the presence of non-constant intrinsic variance brings additional challenges to the analysis; see \citep{Baker2022} for a thorough discussion. Replication, which involves repeating simulation experiments with identical inputs but with varied outputs, offers an effective way to understand intrinsic variations. Additionally, replication has the potential to reduce computational costs for building emulators by averaging repeated observations. Stochastic kriging (SK) is a commonly used emulation technique to approximate the relationship between the inputs and the corresponding stochastic simulation outputs \citep{Ankenman2009}. One drawback of SK is that a second, separate GP model and a minimum level of replication are required to predict the intrinsic variance at any design point. In contrast to the SK approach of separate variance calculations, \cite{Binois2018} propose heteroskedastic GP (hetGP) emulators that jointly infer all unknowns without requiring a minimum level of replication. Although there are other ways of inferring intrinsic variance (see examples in \cite{Kersting2007, Gredilla2011}), in this work, we follow the implementation of hetGP for inference thanks to the accessible software. 

Emulators for stochastic simulation models often utilize identical designs to those employed in deterministic simulations. A common choice is to use a space-filling approach to allocate unique input locations and a uniform number of replicates on each distinct location. A limitation of such designs for calibration is the potential inadequacy in exploring the (unknown) parameter region of interest, particularly in high-dimensional spaces, since the design is constructed without considering the alignment between simulation output and observed data. Moreover, one should carefully determine the number of replicates to better learn the signal and noise relationship varying throughout the parameter region of interest. An experimental design for calibration inference, the topic of this article, comprises the locations of unique parameters and the corresponding replications necessary for the stochastic simulation models. Sequential designs, also known as active learning in the machine learning literature, are often more advantageous compared to the aforementioned one-shot approaches where all design points are chosen in advance. An acquisition function in sequential design quantifies the value of evaluating the simulation model using a specific input, thereby guiding the sequential addition of new simulation data points. Sequential designs enhance budget allocation efficiency by leveraging information gained from prior simulation runs, enabling a more prudent allocation of the remaining simulation budget. Sequential designs can be implemented in two ways: batch sequential, where multiple runs are added simultaneously, or fully sequential, where runs are introduced individually, one at a time. 

Sequential designs have been used to build globally accurate emulators for stochastic simulation models. The integrated mean squared prediction error (IMSE) is one of the most common criteria for general-purpose emulators. For example, \cite{Ankenman2009} introduce a two-stage design approach for constructing accurate SK models. In the first stage, a fixed number of replicates are allocated to a design from LHS, while the optimal number of replicates is determined in the second stage to minimize the IMSE. \cite{ChenZhou2015, Chen2017} propose IMSE-based sequential design strategies to balance exploration over the design space and exploitation with additional replicates using SK. At each active learning step, a fixed budget of additional simulations is allowed, and a decision is made regarding the distribution of replications among existing design points or assigning all replicates to a new design point. \cite{Binois2019} propose a sequential design with hetGP emulators by acquiring one simulation at a time. At each iteration, a decision is made to select either a replicate or a new location to minimize the IMSE. Modern multi-core computing environments allow for the evaluation of the simulation model with a batch of inputs in parallel, and one-at-a-time updates would be inefficient as they fail to leverage all available computational resources. Successful examples of batch sequential procedures for building global emulators of deterministic and stochastic simulation models can be found in \cite{Loeppky2010, Anton2020, Collin2021, Zhang2022}. 

While prior studies have demonstrated the efficiency of active learning combined with an emulator of stochastic models in achieving globally high prediction accuracy, there remains a dearth of research concerning criteria in the calibration context where global prediction is unnecessary. In a recent study by \cite{Surer2023}, the expected integrated variance criterion is introduced for enhancing posterior predictions of parameters in deterministic simulation models. They achieve this by sequentially acquiring new parameters and their high-dimensional outputs. In alignment with this approach, we introduce a batch sequential experimental design to better learn the posterior. At each stage, a batch of simulation evaluations is allocated to existing parameters or new ones utilizing our novel acquisition functions designed for replication and exploration. The proposed acquisition functions aim to minimize the aggregated uncertainty of the posterior, rather than concentrating solely on the overall emulator uncertainty. Our replication selection criterion prioritizes existing parameters from regions with high posterior probability, ensuring greater focus on areas relevant to the calibration of interest. Nonetheless, if the parameters included in the existing simulation data set do not cover the parameter region of interest, the majority of the replicates would likely end up in regions with near-zero posterior density. To understand the behavior of the simulation model near the region of interest, we propose another acquisition function for exploration to select new parameters that minimize the overall uncertainty in the estimate of the posterior density. While allocating more replicates to regions with both high noise levels and posterior probability, our design avoids regions far from the parameter region of interest, even in the presence of high noise levels. Unlike IMSE-based strategies, which may indiscriminately prioritize regions with high noise, our criteria ensure alignment with observed data. In this way, the proposed procedure leads to the better utilization of computational resources and improves calibration efficiency. 

The remainder of the paper is structured as follows. 
Section~\ref{sec:review} outlines the steps of the batch sequential design procedure and provides background information on calibration and emulation. Our methodological contributions are outlined in Section~\ref{sec:acquisition}. We then demonstrate the effectiveness of the proposed approaches in Section~\ref{sec:experiment}, where we analyze results from various simulation experiments, including synthetic and epidemiologic simulation models. Finally, our conclusions are summarized in Section~\ref{sec:conc}.

\section{Review}
\label{sec:review}

\subsection{Batch Sequential Experiment Design}

We consider a system where the real data is observed at design points $\xb_1, \ldots, \xb_d$ $\in \mathbb{R}^{q}$ from a physical experiment. We use $\tg = (\vartheta_1, \ldots, \vartheta_p)^\top$ to denote the vector of generic input parameters to the simulation model, which belongs to the space $\Theta \subset \mathbb{R}^{p}$. Our analysis assumes that simulation outputs are obtained at all design points $\xb_1, \ldots, \xb_d$ once the simulation model, denoted as $\simout$, is evaluated with parameter $\tg$. Specifically, the simulation model takes parameter $\tg$ as an input and produces a vector-valued output $\simout\left(\tg\right) = \left(\zeta\left(\xb_1, \tg\right), \ldots, \zeta\left(\xb_d, \tg\right)\right)^\top$ at $d$ locations where real data is collected. In this setting, the calibration parameter $\tg$ is the only driving input to the simulation model. The design points are not user-defined inputs to the simulation model but are used for notational convenience to represent locations where simulation outputs correspond to observed real data. For the stochastic simulation model, the output is represented as
\begin{equation}
    \simout(\tg) = \latent(\tg) + \nuv, \quad \nuv \sim {\rm MVN}(\mathbf{0}, \Rb(\tg)),  \label{eq:simmodel}
\end{equation}
where $\latent(\tg)$ denotes the expected value, $\mathbb{E}\left[\simout(\tg)\right]$, of the simulation output. The simulation noise, represented by $\nuv$, is a multivariate normal (MVN) random variable with mean zero and covariance matrix $\Rb(\tg)$. The $i$th diagonal element of $\Rb(\tg)$ represents the intrinsic variance of the simulation noise at the design point $\xb_i$ and the $(i, j)$th element of $\Rb(\tg)$ is the covariance between the $(i, j)$th output.

\begin{algorithm}[h]  
    \textbf{Input:} An initial $n_0$ unique parameter locations $\tg_i$, $i=1, \ldots, n_0$, $a_i$ replicates observed at $\tg_i$, a simulation model $\simout(\cdot)$

    \emph{Initialize} $\mathcal{D}_1 = \{(\tg_i, \simout(\tg_i)^l) : l = 1, \ldots, a_i, i = 1, \ldots, n_0\}$
    
    \For {$t = 1,\ldots,T_b$} {
        \emph{Fit} an emulator with $\mathcal{D}_{t}$
        
        \emph{Acquire} a batch of $b$ parameters 

        \emph{Evaluate} the simulation model in parallel with $b$ parameters 

        \emph{Update} $\mathcal{D}_{t+1}$ with $b$ parameters and their simulation outputs
        }

    \textbf{Output:} Simulation data set $\mathcal{D}_{T_b+1}$ along with the emulator fitted with $\mathcal{D}_{T_b+1}$
    
    \caption{Batch sequential experiment design}
    \label{alg:oaat}
\end{algorithm}
Algorithm~\ref{alg:oaat} presents an overview of the batch sequential experimental design procedure. Throughout the paper, we use a subscript index with $\tg$, such as $\tg_i$, to denote the parameters included in the design.
Let $\tg_1, \ldots, \tg_{n_t}$ represent $n_t$ distinct parameter locations collected by stage $t$. Suppose that the simulation model is executed for $a_i$ replicates at each $\tg_i$, resulting in outputs $\simout(\tg_i)^l$ for $l = 1, \ldots, a_i$. The procedure starts by evaluating the simulation model using an initial experimental design comprising a set of $n_0$ unique parameter locations, each with $a_i$ replicates (for $i = 1, \ldots, n_0$). Initial sampling for the unique locations can be carried out using either LHS or randomly from the prior distribution.
During each stage $t$, a batch of $b$ parameters is acquired for evaluation with the simulation model (see line~5). We propose two novel acquisition functions to decide whether a batch of simulation evaluations is assigned to either existing parameter locations or unexplored regions at each stage (see Section~\ref{sec:acquisition}). We consider a batch synchronous procedure where a fixed set of 
$b$ workers are available to perform $b$ simulation evaluations simultaneously.
Once these $b$ evaluations are completed at the end of stage $t$, the simulation data set $\mathcal{D}_{t+1} = \{(\tg_{i}, \simout(\tg_{i})^l): l = 1, \ldots, a_i, i = 1, \ldots, n_t\}$ is updated to include all the simulation data.
The simulation data set, $\mathcal{D}_{t}$, is used to build the emulator (see Section~\ref{sec:GP}), and then the acquisition functions as in lines~4--5. This iterative process continues through $T_b$ batches, leveraging parallel processing and adaptive learning to predict the posterior accurately.

\subsection{Calibration}
\label{sec:notation}

We consider Bayesian calibration, which aims to determine the unknown calibration parameters using data $\yb = (y(\xb_1), \ldots, y(\xb_d))^\top$ from physical experiments and to quantify the uncertainties in the inferred parameters and their corresponding predictions. We refer to $\tb = (\theta_1, \ldots, \theta_p)^\top \in \Theta$ as the unknown calibration parameters that align the simulation model with the data $\yb$. We formulate the data from the physical experiment using the following statistical model
\begin{equation}
    \yb = \latent(\tb) + \epsilonv, \quad  \epsilonv \sim {\rm MVN}\left(\mathbf{0}, \Sigmav\right), \label{eq:statmodel}
\end{equation}
where $\epsilonv$ denotes the residual error.
Alternatively, an additional discrepancy term can be considered to model the observed data (KOH, \cite{Ohagan2001}). While incorporating the discrepancy term can lead to identifiability issues \citep{Higdon2004, Bayarri2007, Jenny2014, Tuo2015, Plumlee2017}, the KOH framework remains a widely used method for calibrating simulation models. Extending our proposed acquisition functions for stochastic models to account for the discrepancy term is left as future work to address this complexity comprehensively. 

According to Bayes' rule, the posterior density $p(\tb|\yb)$ is expressed as follows
\begin{equation} 
\label{eq:posterior}
    p\left(\tb|\yb\right) = \frac{ p\left(\yb|\tb\right) p\left(\tb\right)}{\int_{\Theta}  p\left(\yb|\tb^\prime\right) p\left(\tb^\prime\right) {\rm d \tb^\prime}} \propto \tilde{p}\left(\tb|\yb\right) =  p\left(\yb|\tb\right) p\left(\tb\right),
\end{equation}
where $p(\yb|\tb)$ denotes the likelihood function, indicating how well the simulation output at 
$\tb$ matches the observed data $\yb$. 
The prior probability density $p(\tb)$ reflects initial knowledge about parameter $\tb$, typically expressed as a known, closed-form function. 
The term $\tilde{p}\left(\tb|\yb\right)$ represents the unnormalized posterior. 
According to the model in \eqref{eq:statmodel}, the likelihood is given by
\begin{equation}
   \p\left(\yb|\tb\right) = (2 \pi)^{-d/2} |\Sigmav|^{-1/2} \exp\left(-\frac{1}{2} \left(\yb - \latent\left(\tb\right)\right)^\top \Sigmav^{-1} \left(\yb - \latent\left(\tb\right)\right)\right). \label{eq:truelike}
\end{equation}

Markov chain Monte Carlo (MCMC) methods are commonly employed in Bayesian calibration to generate samples from the posterior distribution \citep{Gilks1995}. Since the normalizing term $\int_{\Theta} p\left(\yb|\tb^\prime\right) p\left(\tb^\prime\right) {\rm d \tb^\prime}$ in the denominator in \eqref{eq:posterior} is independent of $\tb$ and computationally difficult to compute, the unnormalized posterior $\tilde{p}\left(\tb|\yb\right)$ is typically utilized in MCMC frameworks to approximate the posterior up to a constant factor. In this work, while our goal is to learn the shape of the posterior, we do not explicitly construct an estimator for the normalized posterior. Instead, following approaches similar to \cite{Kandasamy2015, Kandasamy2017, Jarvenpa2019, Jarvenpa2021, Surer2023}, we treat the unnormalized posterior as the quantity of interest and focus on assessing the uncertainty associated with its estimation. An alternative approach could involve defining the aggregated uncertainty over the normalized posterior density function. However, such an approach would require discretization of the $\Theta$-space, which poses significant computational and theoretical challenges. To ensure tractability, we target parameter regions with high unnormalized posterior values, as these regions primarily determine the overall shape of the posterior. For simplicity, we will refer to the unnormalized posterior as the posterior throughout the remainder of this paper.

\subsection{Emulator for the High-Dimensional Simulation Output}
\label{sec:GP}

We need an emulator to predict the unknown true mean response $\latent\left(\cdot \right)$ for approximating \eqref{eq:truelike} at any parameter. To achieve this, we treat each design point $\xb_j$ independently for $j = 1, \ldots, d$, by constructing a separate SK model for each observed data point. Consequently, instead of creating a single large emulator for the entire input space, we build $d$ emulators within the $p$-dimensional space. This approach allows us to treat the high-dimensional output as a collection of independent scalar outputs, simplifying the modeling process.  Despite not accounting for the correlation between different outputs, the idea of decomposing a large surrogate model into multiple smaller ones has yielded successful results, as demonstrated in \cite{Huang2020}.  In the following, we introduce some notations and provide a general overview of the SK predictions. We derive the proposed acquisition functions using an independent SK for each design point, assuming known intrinsic variance. Further details on inference and implementation via hetGP are discussed in Section~\ref{sec:inference}. Moreover, Supplementary Material~\ref{sec:emuperformance} further justifies the accuracy of the emulator using real-data experiments from epidemiology.

Consider the unique parameters $\tg_1, \ldots, \tg_{n_t}$ at which the simulation model has been evaluated by the end of stage $t$, and the vector $\ab_t = (a_1, \ldots, a_{n_t})^\top$, which stores the number of replicates at each unique parameter. The sample average simulation output at $\xb_j$ is given by $\bar{\simout}_{t,j} = \left(\frac{\sum\limits_{l=1}^{a_1}\zeta(\xb_j, \tg_1)^l}{a_1}, \ldots, \frac{\sum\limits_{l=1}^{a_{n_t}}\zeta(\xb_j, \tg_{n_t})^l}{a_{n_t}}\right)^\top$ across $\ab_t$ replicates for $j = 1, \ldots, d$. 
The kernel function $k_{t,j}(\cdot, \cdot) = \tau_{t,j} c_{t,j}(\cdot, \cdot)$ specifies the covariance structure between two parameters with the scaling parameter $\tau_{t,j}$ and the correlation function $c_{t,j}(\cdot, \cdot)$. 
Due to our approach of constructing independent emulators at each $\xb_j$, we approximate the covariance matrix $\Rb(\tg)$ in \eqref{eq:simmodel} using a diagonal matrix of size $d \times d$, where the diagonal elements are denoted as $r_j(\tg)$ for $j = 1, \ldots, d$. Here, $r_j(\tg)$ denotes the variance of the intrinsic uncertainty inherent in the stochastic simulation output $j$, which varies with the input parameter $\tg$.  Moreover, $r_{j}(\tg)$ has no relation to $r_{j^\prime}(\tg)$ for two distinct design points $\xb_{j}$ and $\xb_{j^\prime}$ at any given parameter $\tg$, where $1 \leq j, j^\prime \leq d$ and $j \neq j^\prime$.  Let $\Kb_{t,j}$ denote the $n_t \times n_t$ covariance matrix, where the $(i, i')$th entry is given by $k_{t,j}(\tg_i,\tg_{i'})$ for $1 \leq i, i' \leq n_t$. According to the GP prior, $\latent_{t,j} = \left(\eta(\xb_j, \tg_1), \ldots, \eta(\xb_j, \tg_{n_t})\right)^\top$ is a multivariate normal random vector with mean $\mathbf{0}$ and $n_t \times n_t$ covariance matrix $\Kb_{t,j}$ such that $\latent_{t,j}\sim\text{MVN}(\mathbf{0}, \Kb_{t,j})$. Throughout this text, we distinguish between intrinsic uncertainty, represented by $r_j(\cdot)$, and extrinsic uncertainty, modeled by the kernel function $k_{t,j}(\cdot, \cdot)$, which captures correlations among latent variables $\eta(\xb_j, \cdot)$ \citep{Ankenman2009}.
Leveraging the conditional properties of MVN distributions, SK predictions are characterized by the mean $m_{t,j}(\tg)$ and variance $\varsigma^2_{t,j}(\tg)$ as follows
\begin{equation}
    \begin{aligned}
        & m_{t,j}(\tg) = \kb_{t,j}(\tg)^\top \Kb_{t,j}(\ab_t)^{-1} \bar{\simout}_{t,j} \text{ and }  \varsigma^2_{t,j}(\tg) = k_{t,j}(\tg, \tg) - \kb_{t,j}(\tg)^\top \Kb_{t,j}(\ab_t)^{-1} \kb_{t,j}(\tg), \label{eq:meanvar_latent}
    \end{aligned}
\end{equation}
\begin{gather}
   \text{where} \quad \Kb_{t,j}(\ab_t) = \Kb_{t,j} + \Vb_{j}(\ab_t) \quad \text{and} \quad \Vb_{j}(\ab_t) = \text{diag}\left(\frac{r_j(\tg_1)}{a_1}, \ldots, \frac{r_j(\tg_{n_t})}{a_{n_t}}\right). \notag
\end{gather}
The cross-kernel evaluations between $\tg$ and $\tg_{i}$, $i=1, \ldots, n_t$, is stored in $\kb_{t,j}(\tg) = (k_{t,j}\left(\tg, \tg_1\right),$ $ \ldots, k_{t,j}\left(\tg, \tg_{n_t}\right))^\top$. Letting $\muv_t(\tg) = (m_{t,1}(\tg), \ldots, m_{t,d}(\tg))^\top$ and $\Sv_t(\tg) = \operatorname{diag}\left(\varsigma^2_{t,1}(\tg), \ldots, \right. \allowbreak \left. \varsigma^2_{t,d}(\tg)\right)$, the predictive distribution of the emulator output is
\begin{equation}
    \latent(\tg)|\mathcal{D}_{t} \sim \text{MVN}\left(\muv_t(\tg), \Sv_t(\tg)\right),
    \label{emu_final}
\end{equation}
where $\muv_t(\tg)$ is the emulator predictive mean and $\Sv_t(\tg)$ is the covariance matrix.

\section{Integrated Variance for Calibration of Stochastic Models}
\label{sec:acquisition}

We propose two novel acquisition functions: one selects a batch of $b$ replicates allocated to existing unique locations, while the other targets parameters from unexplored regions to improve posterior learning. At each stage of Algorithm~\ref{alg:oaat}, we decide whether to allocate $b$ simulation runs for replication or exploration to minimize the overall posterior uncertainty. Alternatively, one could consider combining replication and exploration within the same batch. However, this approach requires carefully designed decision rules to effectively balance the proportion of runs allocated to each, which must be flexible enough to work across a variety of applications and parallel resource configurations. To simplify the decision-making process, we adopt a strategy inspired by \cite{Chen2017}, where resources within a batch are allocated to either existing design points or new ones.

The acquisition functions for replication and exploration are detailed in Sections~\ref{sec:exploitation}--\ref{sec:exploration}, respectively. In Lemma~\ref{lemma:UQ}, the expectation $\E[\tilde{p}(\tb|\yb)]$ serves as the estimate of the posterior density, while the variance $\V[\tilde{p}(\tb|\yb)]$ quantifies the uncertainty in the posterior at $\tb$. Both are used to construct the proposed acquisition functions discussed in the following sections. The proof of Lemma~\ref{lemma:UQ} is derived from Lemma~3.1 in \cite{Surer2023} by substituting the deterministic simulation model output with the expected value of the output from the stochastic simulation model.
    \begin{lemma}\label{lemma:UQ}
    Under \eqref{eq:statmodel}, \eqref{eq:posterior}, and \eqref{emu_final}, 
            \begin{gather}
                 \E\left[\tilde{p}\left(\tb|\yb\right)\right] = f_\mathcal{N}\left(\yb; \muv_t\left(\tb\right), \Sigmav + \Sv_t\left(\tb\right)\right) p\left(\tb\right),  \label{expectedpostfinal} \\
                 \V\left[\tilde{p}\left(\tb|\yb\right)\right] = \left(\frac{f_\mathcal{N}\left(\yb; \muv_t\left(\tb\right), \frac{1}{2}\Sigmav + \Sv_t\left(\tb\right)\right)}{2^d\pi^{d/2}|\Sigmav|^{1/2}}  - \left(f_\mathcal{N}\left(\yb; \muv_t\left(\tb\right),\Sigmav + \Sv_t\left(\tb\right)\right)\right)^2\right)p\left(\tb\right)^2. 
             \end{gather}
    \end{lemma}

In the paper, $f_\mathcal{N}(\mathbf{a}; \mathbf{b}, \mathbf{C})$ represents the probability density function of the normal distribution with mean $\mathbf{b}$ and covariance $\mathbf{C}$, evaluated at the value $\mathbf{a}$.
\begin{figure}[t]
\centering
    \begin{subfigure}{1\textwidth}    
        \includegraphics[width=1\textwidth]{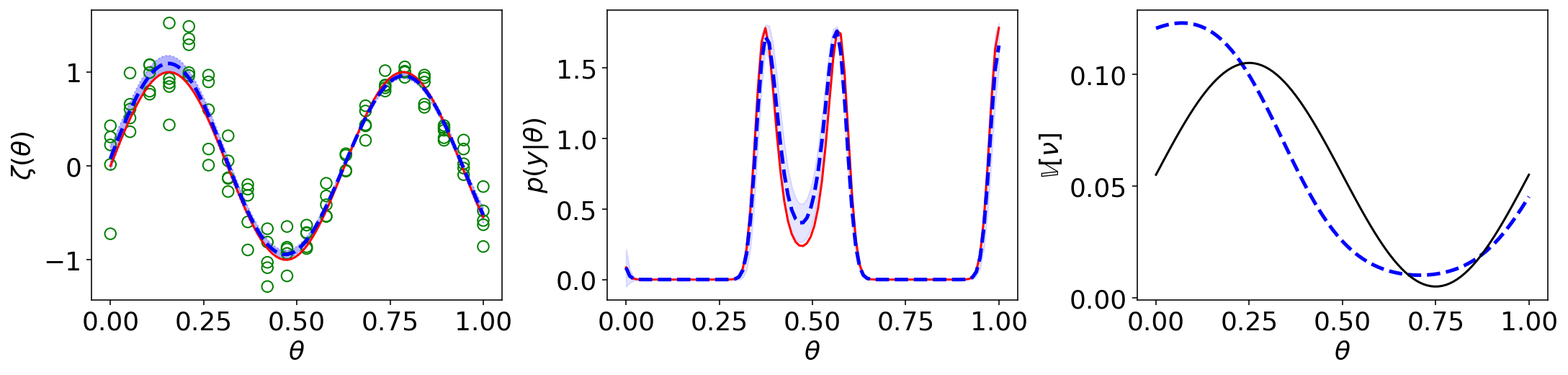}
    \end{subfigure}
    \caption{Illustration with a simulation model $\zeta(\vartheta) = \eta(\vartheta) + \nu$ where $\eta(\vartheta) = \sin(10\vartheta)$ and $\nu \sim {\rm N}(0, 1.1 + 0.05 \sin(2\pi\vartheta))$. The observed data is generated through $y = \eta(\theta = 0.5) + \epsilon$ with $\epsilon \sim {\rm N}(0, 0.05^2)$. The red line shows the expected value of the simulation model (left panel) or the likelihood (middle panel). The blue dashed line shows the prediction mean and the shaded area illustrates one predictive standard deviation from the mean. Green dots indicate the simulation data including five replicates of 20 uniformly spaced parameter values used to build the emulator. The right panel demonstrates the true (black line) and estimated (blue dashed line) intrinsic variance.}
    \label{fig:onedexample}
\end{figure}

To demonstrate proposed acquisition functions, we consider the stochastic simulation model $\zeta(\vartheta)$ with parameter $\vartheta$, as shown in Figure~\ref{fig:onedexample}. Boldface characters are omitted from $\eta(\vartheta)$, $\zeta(\vartheta)$, $y$, and $\vartheta$ since they represent scalars. The left panel displays the emulator constructed using simulation data, consisting of 20 unique parameter values, each replicated five times. The likelihood prediction and its variance in the middle panel are computed using Lemma~\ref{lemma:UQ} with the emulator. We note that, in this example, the posterior is entirely determined by the likelihood because we assume a uniform prior. The noise function shown in the right panel mirrors the one-dimensional example introduced in \cite{Binois2019}. In the following sections, we outline the proposed acquisition functions and showcase their application using this illustrative example.

\subsection{Integrated Variance for Replication}
\label{sec:exploitation}

Recall that by the end of stage $t$, $\ab_t = \left(a_1, \ldots, a_{n_t}\right)^\top$ denotes the replications that have been allocated to distinct parameters $\tg_1, \ldots, \tg_{n_t}$. In the next stage, we have $n_t$ parameters to which we consider allocating $\Delta \ab_t = \left(\Delta a_1, \ldots, \Delta a_{n_t}\right)^\top$ replicates such that $\sum\limits_{i=1}^{n_t} \Delta a_{i} = b$. The IMSE criterion, introduced by \cite{Ankenman2009} in a two-stage procedure and extended by \cite{Chen2017} in a sequential procedure, aims to determine $\Delta \ab_t$ that minimizes the aggregated uncertainty of the emulator. Inspired by this approach, we also consider total uncertainty, but in a novel way: we focus on the aggregated variance of the posterior across the entire parameter space to better learn the shape of the posterior.

Let $\DatExtR$ denote the extended simulation data set, which includes the data $\mathcal{D}_t$ collected up to stage $t$ and the hypothetical simulation data set resulting from allocating $\Delta \ab_t$ replicates. Formally, it is defined as $\DatExtR = \mathcal{D}_t \cup \left(\left(\tg_i, \simout(\tg_i)^l\right) : l = 1, \ldots, \Delta a_i, i = 1, \ldots, n_t\right)$. To derive the optimal allocation of replications, we minimize the integrated variance (IVAR) such that 
\begin{align} \notag 
    \begin{split}
        {\rm IVAR}(\Delta \ab_t) &= \int\limits_{\tb \in \Theta} \VR\left[\tilde{p}\left(\tb|\yb\right)\right] d\tb = \int\limits_{\tb \in \Theta}\left(\frac{f\left(\tb\right)}{2^d\pi^{d/2}|\Sigmav|^{1/2}}  - g\left(\tb\right)^2\right)p\left(\tb\right)^2 d\tb, \\ 
         \text{where} \quad f(\tb) &= f_\mathcal{N}\left(\yb; \muv_{t}^\Delta(\tb), \frac{1}{2}\Sigmav + \Sv_{t}^\Delta(\tb)\right) \quad \text{and} \quad g(\tb) = f_\mathcal{N}\left(\yb; \muv_{t}^\Delta(\tb),\Sigmav + \Sv_{t}^\Delta(\tb)\right).
    \end{split}
\end{align}
Here, $\muv_{t}^\Delta(\tb) = (m_{t,1}(\tb, \Delta \ab_t), \ldots, m_{t,d}(\tb, \Delta \ab_t))^\top$ denotes the vector of updated mean predictions at $\tb$, while $\Sv_{t}^\Delta\left(\tb\right) = {\rm diag}\left(\varsigma^2_{t,1}(\tb, \Delta \ab_t), \ldots, \varsigma^2_{t,d}(\tb, \Delta \ab_t)\right)$ is the diagonal covariance matrix, with each diagonal entry representing the variance of the updated predictions at a design point. Since no new simulation outputs are observed when considering $\Delta \ab_t$ allocations, the emulator's hyperparameters remain fixed during the computation of ${\rm IVAR}(\Delta \ab_t)$. To obtain $m_{t,j}(\tb, \Delta \ab_t)$ and $\varsigma^2_{t,j}(\tb, \Delta \ab_t)$, the diagonal elements of $\Kb_{t,j}(\ab_t)$ in \eqref{eq:meanvar_latent} (i.e., $\Vb_{j}(\ab_t)$) must be modified to reflect $\ab_t + \Delta \ab_t$ replicates. Furthermore, to account for the sample average of the unseen outputs, the existing sample averages $\bar{\simout}_{t,j}$ in \eqref{eq:meanvar_latent} must be adjusted to include the contribution of the hypothetical sample averages. Since these unseen sample averages are unknown under the data $\mathcal{D}_t$, one can consider $m_{t,j}(\tb, \Delta \ab_t)$ in expectation. This requires imputing the unknown sample averages at $\tg_1, \ldots, \tg_{n_t}$ using $m_{t,j}(\tg_1), \ldots, m_{t,j}(\tg_{n_t})$. It is observed that the posterior predictions obtained with the updated mean degrade as the batch size $b$ increases. Therefore, to derive the allocations via ${\rm IVAR}(\Delta \ab_t)$, we assume that the emulator mean $\muv_{t}^\Delta\left(\tb\right)$ remains unaffected by the additional $\Delta \ab_t$ replicates, such that $\muv_{t}^\Delta\left(\tb\right) = \muv_{t}\left(\tb\right)$. 

We aim to minimize IVAR as a function of the number of additional replicates allocated to each design point. Following \cite{Ankenman2009}, we formulate this problem as a constrained optimization problem such that
    \begin{subequations} \label{prob:repopt}
        \begin{align}
            \min_{\Delta \ab_t} \quad & {\rm IVAR}(\Delta \ab_t) && \label{eqn:objective_function} 
            \\
            \st \quad & \sum_{i=1}^{n_t} \Delta a_i \leq b, \label{eqn:cons1}  \\
            & \Delta a_i \in \mathbb{Z}^+, \forall i = 1, \ldots, n_t. \label{eqn:cons2} 
        \end{align}
    \end{subequations}
 The optimization problem in \eqref{prob:repopt} seeks the optimal additional $\Delta \ab_t$ replicates that minimize $\allowbreak {\rm IVAR}(\Delta \ab_t)$ as specified in the objective function \eqref{eqn:objective_function}. Constraint~\eqref{eqn:cons1} ensures that the total number of additional replicates does not exceed $b$, while Constraint~\eqref{eqn:cons2} requires each $\Delta a_i$ to take a nonnegative integer value. Model \eqref{prob:repopt} is a nonlinear integer program, which is significantly harder to solve than integer linear programs \citep{Hemmecke2010}. While off-the-shelf solvers can be employed by leveraging the derivative $\frac{\partial  {\rm IVAR}(\Delta \ab_t)}{\partial \Delta a_i}$, as provided in Supplementary Material~\ref{proof:ivar_rep}, solving nonlinear integer programs presents substantial challenges, including the risk of converging to suboptimal solutions and increased computational time as the number of integer variables grows. To avoid the challenges of difficult numerical optimization, we use an approximate formula by relaxing the integrality constraint and considering only $a_i \geq 0$. The following result proves helpful in obtaining an approximate solution to the resulting nonlinear program using its first-order conditions, with the proof provided in Supplementary Material~\ref{proof:ivar_rep}.

\begin{theorem}
Suppose that $\ab_t = \left(a_1, \ldots, a_{n_t}\right)^\top$ replicates have been allocated at $\tg_1, \ldots, \tg_{n_t}$. We consider allocating additional $\Delta \ab_t = \left(\Delta a_1, \ldots, \Delta a_{n_t}\right)^\top$ replicates to these points during stage $t+1$. Let $\Delta \mathbf{V}_j(\ab_t) = {\rm diag}(v_{j,1}, \ldots, v_{j,n_t})$ where $v_{j,i} = \frac{r_j(\tg_i)}{a_i + \Delta a_i}-\frac{r_j(\tg_i)}{a_i}=-\frac{r_j(\tg_i)\Delta a_i}{(a_i + \Delta a_i)a_i}$, $i = 1, \ldots, n_t$, $j = 1, \ldots, d$. If $\max \limits_{i=1, \ldots, n_t} \left[\Delta \mathbf{V}_j(\ab_t)\right]_{i,i} \ll 1$ such that $\mathbf{I} + \Delta \mathbf{V}_j(\ab_t) \Kb_{t,j}(\ab_t)^{-1} \approx \mathbf{I}$, then the optimal budget allocation rule that leads to the maximum reduction in IVAR is to assign $\Delta a_i$ replicates to $\tg_i$ such that
\begin{gather}
    \begin{aligned}
        a_i + \Delta a_i &\propto \sqrt{C_{i}(\ab_t)}, \quad i = 1, \ldots, n_t, \quad \text{where} \label{eq:ivar_allocation} \\
    \end{aligned} \\
        C_{i}(\ab_t) = \int\limits_{\tb \in \Theta} \left( \frac{f(\tb)}{2^d\pi^{d/2}|\Sigmav|^{1/2}}  \left(-\frac{1}{2} {\rm Tr}\left(\dot{\mathbf{N}}(\tb)^{-1} \Mb_i(\tb) \right) + \frac{1}{2}\mathbf{h}(\tb)^\top \dot{\mathbf{N}}(\tb)^{-1} \Mb_i(\tb) \dot{\mathbf{N}}(\tb)^{-1} \mathbf{h}(\tb)\right) \right. \nonumber \\ \left.
        \hspace{1cm} - 2 g(\tb)^2 \left(-\frac{1}{2} {\rm Tr}\left(\mathbf{N}(\tb)^{-1}\Mb_i(\tb) \right) + \frac{1}{2}\mathbf{h}(\tb)^\top \mathbf{N}(\tb)^{-1} \Mb_i(\tb) \mathbf{N}(\tb)^{-1} \mathbf{h}(\tb)\right) \right) p(\tb)^2 d\tb, \nonumber 
\end{gather}
with $\dot{\mathbf{N}}(\tb) = 0.5\Sigmav + \Sv_{t}(\tb)$, $\mathbf{N}(\tb) = \Sigmav + \Sv_{t}(\tb)$, $\mathbf{h}(\tb) = \yb-\muv_{t}(\tb)$, and $\Mb_i(\tb)$ is a $d \times d$ diagonal matrix with $j$th diagonal element $-r_j(\tg_i)\kb^\top_{t,j}(\tb) \Kb_{t,j}(\ab_t)^{-1}  \mathbf{J}^{(i,i)} \Kb_{t,j}(\ab_t)^{-1} \kb_{t,j}(\tb)$. $\mathbf{J}^{(i,i)}$ is a $n_t \times n_t$ matrix with one in the $(i,i)$th entry and zeros elsewhere.
\end{theorem}
During stage $t+1$, the allocation according to \eqref{eq:ivar_allocation} requires each $a_i + \Delta a_i$ to be proportional to $\sqrt{C_{i}(\ab_t)}$, for $i = 1, \ldots, n_t$. The $j$th diagonal element of $\Mb_i(\tb)$ corresponds to the IMSE criterion for output $j$, as detailed in Supplementary Material~\ref{proof:imse_exploit}. This criterion allocates replicates based on a value that combines intrinsic and extrinsic uncertainty to construct globally accurate emulators. Unlike the IMSE criterion, the proposed IVAR criterion not only accounts for the extrinsic correlation structure and intrinsic variance but also prioritizes replicates in parameter regions of interest. In the proposed allocation, $\Mb_i(\tb)$ incorporates the similarity between $\tg_i$ and $\tb$ through extrinsic correlation, scaled by the intrinsic variance $r_j(\tg_i)$. Combined with the remaining terms, this ensures that parameters $\tg_i$, for $i = 1, \ldots, n_t$, which are similar to $\tb$ in high posterior regions, receive greater weight for additional replicates. Consequently, the IVAR replication strategy promotes replicates in high posterior regions with larger extrinsic and intrinsic variances to enhance posterior learning, while avoiding regions with near-zero posterior density, even when variances are high.

The allocation rule \eqref{eq:ivar_allocation} may propose assigning fewer replicates to certain design points than what is already assigned in the current design, a common challenge in both \cite{Ankenman2009} and \cite{Chen2017}. Since the allocations $a_i$ from previous stages are fixed, they cannot be adjusted. Furthermore, these prior allocations may no longer be optimal at stage $t+1$, as they were made without considering potential selections in future stages. To address this issue, we initially consider allocating a total of $\sum\limits_{i=1}^{n_t}a_i + b$ replicates based on \eqref{eq:ivar_allocation}, assuming that a budget of $\sum\limits_{i=1}^{n_t}a_i + b$ simulation evaluations are possible. This allocation sets an upper bound ${ub}_i$, $i = 1, \ldots, n_t$, for each design point. If the upper bound falls below the replicates already assigned (i.e., ${ub}_i < a_i$), it indicates that fewer replicates might have been optimal for those specific design points. In such instances, we adjust the associated upper bound to zero and allocate $b$ replicates by considering modified ${ub}_i$ as the optimal allocation weights. Since the weighted allocations do not guarantee the integrality constraint, we round the replicates to integer values, ensuring that $\sum\limits_{i=1}^{n_t} \Delta a_i$ does not exceed $b$. Additionally, the allocation rule in \eqref{eq:ivar_allocation} involves an integral over a multi-dimensional parameter space, which we approximate with a sum over uniformly distributed reference grid $\Theta_{\rm ref}$. 

\begin{figure}[ht]
\centering
    \begin{subfigure}{1\textwidth}    
        \includegraphics[width=1\textwidth]{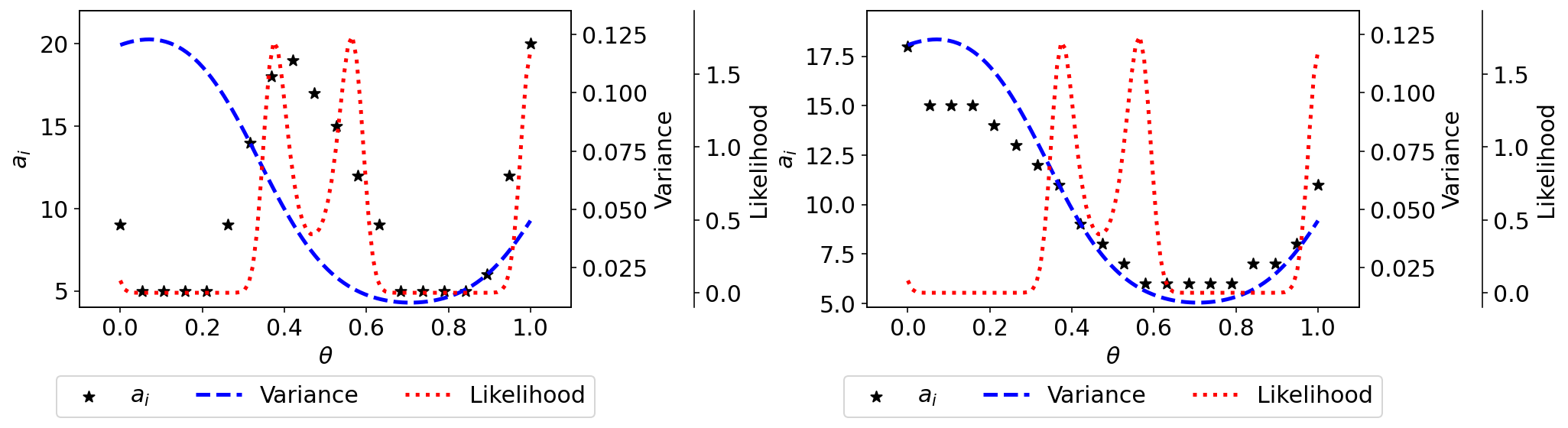}
    \end{subfigure}
    \caption{Allocation of new 100 replicates guided by the proposed acquisition function (left) and \cite{Ankenman2009} (right), using the example illustrated in Figure~\ref{fig:onedexample}. Star markers indicate the number of replicates on the existing 20 design points. The estimated intrinsic variance (blue dashed line) and likelihood (red dotted line) are depicted for reference.}
    \label{fig:allocation_reps}
\end{figure}
Returning to our illustrative example presented in Figure~\ref{fig:onedexample}, the left panel of Figure~\ref{fig:allocation_reps} shows the proposed allocation of an additional 100 replicates ($b=100$) across the existing 20 unique design points ($n_0 = 20$), each initially assigned 5 replicates ($\ab_0 = \mathbf{5}_{20})$. The right panel demonstrates the allocation strategy informed by \cite{Ankenman2009}. In this example, the parameter regions on the left-hand side generally exhibit higher uncertainty due to intrinsic noise in the stochastic simulation outputs compared to those on the right-hand side. As a result, IMSE allocates more replicates to regions with higher intrinsic noise, while fewer replicates are assigned to regions with lower noise levels. However, this approach does not consider aligning simulation outputs with observed data, leading to inadequate targeting of high posterior regions with comparatively lower intrinsic noise. Such allocation, primarily driven by intrinsic variance, may pose challenges, particularly if the parameter region of interest is small relative to prior regions' support, and noise levels vary across the response surface. The proposed approach overcomes this limitation by accounting for both the intrinsic variance of the simulation noise and the extrinsic correlations, while also prioritizing alignment with the observed data. This alignment allows the method to effectively direct replicates to regions that contribute the most to posterior learning and enhance the ability to distinguish signal from noise within the parameter region of interest. Additionally, it avoids selecting additional replicates from the leftmost region, located far from the calibration region of interest, despite its exposure to higher intrinsic noise intensity. Moreover, the proposed strategy prioritizes allocating more replicates from high posterior regions with less uncertainty, such as those in the rightmost region, to better understand the simulation model's behavior near the region of interest. In stage $t=1$, the proposed allocation rule recommends assigning fewer than five replicates to points on the right side--those outside the parameter region of interest with lower uncertainty--even though each already has five replicates. For these points, the upper bound ${ub}_i$ is lower than the current allocation $a_i$. To address this, we set their upper bounds to zero and then allocate the additional $b$ replicates according to the updated ${ub}_i$ values. Consequently, no further replicates are distributed to these points, which have already exceeded their ideal allocation at stage $t=1$.

\subsection{Integrated Variance for Exploration}
\label{sec:exploration}

The replication strategy presented in Section~\ref{sec:exploitation} provides valuable insights into the model's inherent variability within the parameter region of interest. However, without careful selection of unique parameters $\tg_1, \ldots, \tg_{n_t}$, computational resources would be wasted on replicating model evaluations outside this region. Exploration is essential to distribute simulation runs over the calibration region and reduce response-surface uncertainty by guiding the selection of new design points. To facilitate posterior learning, we propose an acquisition function based on the aggregated variance of the posterior over the parameter space for stochastic simulations.

Algorithm~\ref{alg:krigingbeliever} describes the process of constructing a batch of $b$ parameters for exploration. For constructing a batch of size $b$, we consider selecting $\breve{b}$ new points and assigning $\breve{a}$ replications to each new design point such that $b = \breve{b} \times \breve{a}$. The values $\breve{b}$ and $\breve{a}$ are user-specified inputs to our method. A larger $\breve{b}$ facilitates broader exploration of the parameter space, which is essential for identifying high-posterior regions. Conversely, increasing the number of replications $\breve{a}$ enhances the reliability of estimated outputs for new parameters by reducing noise, particularly in areas where the model exhibits high variability. Alternatively, one can consider assigning a batch of $b$ runs to a single new point as in \cite{Chen2017} (i.e., $\breve{b} = 1$ and $\breve{a} = b$). However, with the growing accessibility of many-node servers and large-scale cloud computing environments, we can evaluate the simulation model with a large batch of $b$ parameters in parallel within a sequential design. Therefore, assigning a large batch of $b$ runs to a single new point may not effectively facilitate thorough exploration. To generalize it further, we provide our results with different combinations of $\breve{b}$ and $\breve{a}$. Since choosing $\breve{b}$ locations at once is a more challenging problem than assigning additional $b$ replicates to existing $n_t$ unique points, we use a greedy strategy to iteratively construct a $\breve{b}$-point design.
\begin{algorithm}[H]  
    \textbf{Input:} Simulation data set $\mathcal{D}_t$, an emulator, $\breve{a}$, $\breve{b}$

    \For {$\breve{t} = 1,\ldots,\breve{b}$} {
        \emph{Generate} candidate parameters $\mathcal{L}_{\breve{t}}$
        
        \emph{Select} $\ct^* \in \argmin\limits_{\ct \in \mathcal{L}_{\breve{t}}} {\rm IVAR}(\ct, \breve{a})$
                    
                    
        \emph{Update} the emulator using $\ct^*$ 
        }

    \textbf{Output:} A batch of $\breve{b}$ new parameters, each replicated $\breve{a}$ times
    
    \caption{Acquisition of a batch of $b$ parameters for exploration}
    \label{alg:krigingbeliever}
\end{algorithm}

In the following, we detail this greedy strategy, which builds a $\breve{b}$-parameter design--where each parameter includes $\breve{a}$ replicates--by approximately solving the $p$-dimensional problem $\breve{b}$ times, as outlined in Algorithm~2. At each iteration, we measure the value of replicating the simulation model $\breve{a}$ times at a candidate parameter $\ct$ from the discrete set $\mathcal{L}_{\breve{t}}$ of parameters (see line~4) via the following IVAR criterion 
\begin{align} \label{IVAR_explore}
    \begin{split}
        {\rm IVAR}(\ct, \breve{a}) &= \int\limits_{\tb \in \Theta} \Exp\left(\VE\left[\tilde{p}(\tb|\yb) \right]\right) d\tb,
    \end{split}
\end{align}
where the extended simulation data set $\DatExt$ is given by $\DatExt = \mathcal{D}_t $ $\cup \left(\left(\ct, \simout(\ct)^l\right) : l = 1, \ldots, \breve{a}\right)$. Here, $\simout(\ct)^l$ denotes the $l$th unknown simulation output at $\ct$. Additionally, $\bar{\breve{\simout}} = \left(\bar{\breve{\zeta}}_1, \ldots, \bar{\breve{\zeta}}_d\right)^\top $ represents the hypothetical sample mean of simulation outputs across $\breve{a}$ replicates, with $\bar{\breve{\zeta}}_j = \frac{\sum\limits_{l=1}^{\breve{a}}\zeta(\xb_j, \ct)^l}{\breve{a}}$ for $j = 1, \ldots, d$. The expectation is taken over $\bar{\breve{\simout}}$, which remains random because it is unknown under data $\mathcal{D}_{t}$.

In line~4 of Algorithm~\ref{alg:krigingbeliever}, the next best parameter $\ct^*$ to include in the $\breve{b}$-point design is chosen to minimize the overall posterior uncertainty as if its $\breve{a}$ unknown outputs are included into the simulation data set. 
Since the noisy simulation outputs for the selected parameter are not available during batch construction, the emulator's hyperparameters remain fixed. In line~5, we update $\Kb_{t,j}(\ab_t)$ using \eqref{updatedK} in Supplementary Material~\ref{proof:imse} (with $\ct$ replaced by $\ct^*$). This update requires an estimate of $r_j(\ct^*)$, which is predicted using the emulator as described in Section~\ref{sec:inference}. Additionally, because the sample average simulation output at $\ct^*$ across $\breve{a}$ replicates is not yet available, as noted in \eqref{mean_future} in Supplementary Material~\ref{proof:lemma3.3}, $\bar{\simout}_{t,j}$ is augmented with its expectation, $m_{t,j}(\ct^*)$.
When forming a batch for deterministic models in Bayesian optimization (BO), treating ordinary kriging predictions as true values for subsequent iterations is known as the kriging believer (KB) strategy \citep{Ginsbourger2010}.  While the KB strategy has been criticized for its limited exploration in BO, the risk of getting trapped in suboptimal regions is mitigated in our approach due to the design of our acquisition function. The IVAR criterion is specifically tailored to learn the overall shape of the posterior, thereby inherently promoting exploration. It encourages sampling from regions with low, medium, and high posterior density, rather than focusing solely on the maximum posterior region. However, it is important to note that this approach can introduce prediction inaccuracies, as it relies on the emulator’s estimates for unseen outputs. These inaccuracies may be particularly pronounced when the emulator fails to accurately capture the true model behavior in unexplored regions. In such cases, alternative batch acquisition strategies could be considered, each presenting distinct trade-offs between exploration and replication. The next result is used to compute the ${\rm IVAR}(\ct, \breve{a})$ criterion, and the derivation is provided in Supplementary Material~\ref{proof:lemma3.3}. \\ 

\begin{lemma}\label{prop:IEV}
Let $\PHI_t(\tb, \ct)$ be the $d \times d$ diagonal matrix with diagonal elements $\frac{\text{cov}_{t,j}(\tb, \ct)^2}{\varsigma^2_{t,j}(\ct) + r_j(\ct)/\breve{a}}$.
${\rm IVAR}(\ct, \breve{a})$ is computed via
    \begin{gather} 
        \int\limits_{\tb \in \Theta} p(\tb)^2\left(\frac{f_\mathcal{N}\left(\yb; \, \muv_t(\tb), \, \frac{1}{2}\Sigmav + \Sv_t(\tb)\right)}{2^d \pi^{d/2} |\Sigmav|^{1/2}} -  \frac{f_\mathcal{N}\left(\yb; \, \muv_t(\tb), \, \frac{1}{2}\left(\Sigmav + \Sv_t(\tb) + \PHI_{t}(\tb, \ct)\right)\right)}{2^d \pi^{d/2} \left|\Sigmav + \Sv_t(\tb) - \PHI_{t}(\tb, \ct)\right|^{1/2}}\right)d\tb. \label{eq:IVARexplore}
    \end{gather}
\end{lemma}
To better understand this result, consider a scenario where the total number of replicates, $\sum\limits_{i=1}^{n_t} a_i$, is sufficiently large. In such cases, we can approximate $\Kb_{t,j}(\ab_t) \approx \Kb_{t,j}$ and $\bar{\simout}_{t,j} \approx \latent_{t,j}$ in \eqref{eq:meanvar_latent}. Consequently, the goal of strategically selecting parameters by minimizing total posterior uncertainty is to closely approximate the true posterior while reducing the predictive variance toward zero.
Since integrating the posterior variance over a multi-dimensional parameter space is analytically difficult, we approximate \eqref{eq:IVARexplore} by summing over the discrete set of points $\Theta_{\rm ref}$, similar to the acquisition rule for replication. Additionally, we drop the first term in \eqref{eq:IVARexplore}, since it does not depend on $\ct$. As a result, minimizing \eqref{eq:IVARexplore} for any candidate $\ct$ is approximated by maximizing the following expression:
    \begin{gather} 
        \frac{1}{\Theta_{\rm ref}}\sum_{\tb \in \Theta_{\rm ref}} p(\tb)^2 \left(\frac{f_\mathcal{N}\left(\yb; \, \muv_t(\tb), \, \frac{1}{2}\left(\Sigmav + \Sv_t(\tb) + \PHI_{t}(\tb, \ct)\right)\right)}{2^d \pi^{d/2} \left|\Sigmav + \Sv_t(\tb) - \PHI_{t}(\tb, \ct)\right|^{1/2}}\right).\label{eq:IVARexplore_app}
    \end{gather}

\begin{figure}[t]
\centering
    \begin{subfigure}{1\textwidth}    
        \includegraphics[width=1\textwidth]{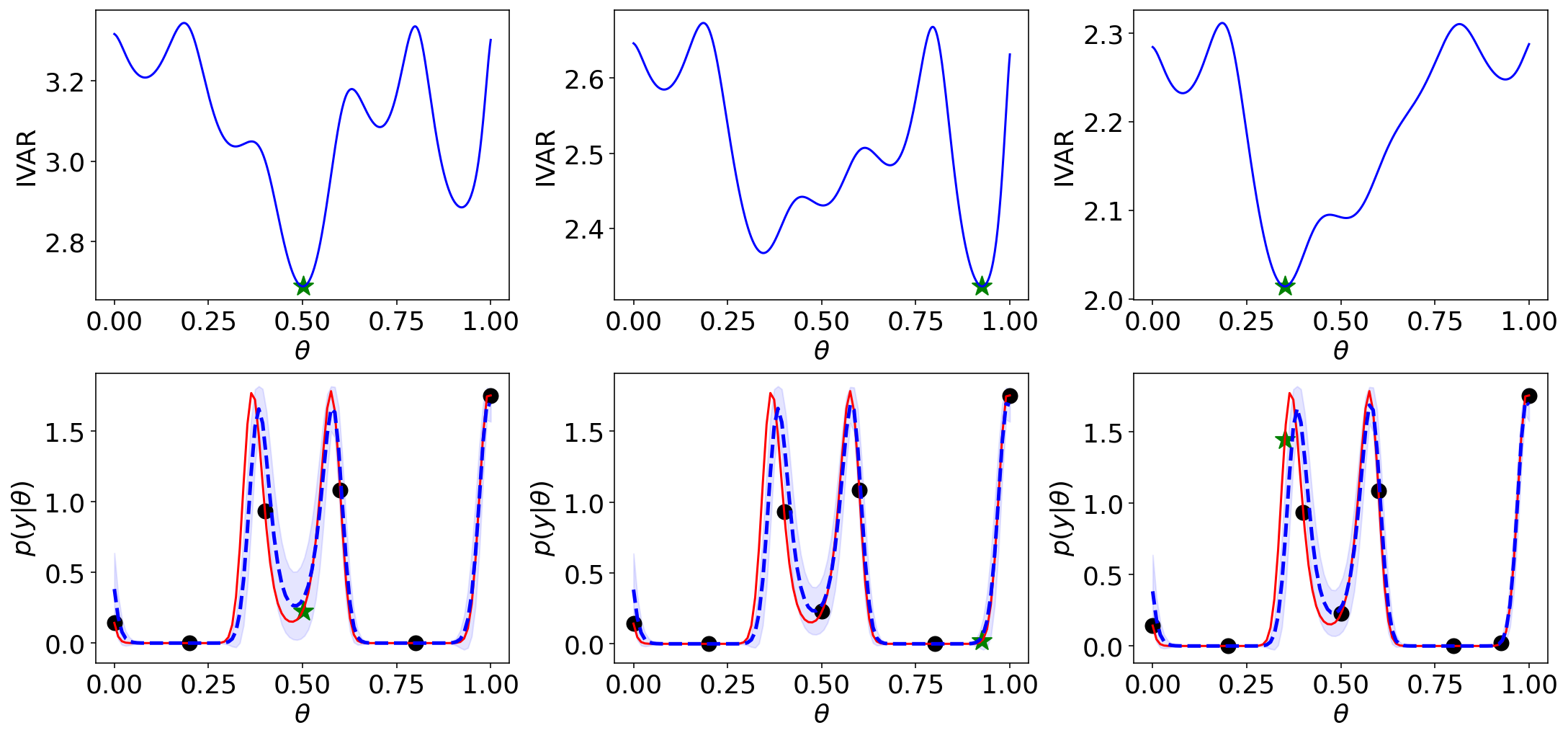}
    \end{subfigure}
    \caption{Allocation of $b=15$ simulation evaluations guided by the IVAR criterion shown in the first row with $\breve{b} = 3$ and $\breve{a}=5$, using the model illustrated in Figure~\ref{fig:onedexample}. The emulator is constructed using simulation data including $n_0 = 6$ unique parameters (black dots), each replicated five times. The green star represents the acquired point. The second row shows the true (red line) and estimated likelihood (blue dashed line).}
    \label{fig:allocation_exp}
\end{figure}

Revisiting the illustrative example in Figure~\ref{fig:onedexample}, Figure~\ref{fig:allocation_exp} displays three iterations of Algorithm~\ref{alg:krigingbeliever} to select $\breve{b}=3$ new points and allocate $\breve{a}=5$ replicates within a batch of $b=15$. The emulator is built with the simulation data set from six evenly spaced points on the interval $[0, 1]$, each with five replicates. The ${\rm IVAR}(\breve{\vartheta}, \breve{a})$ criterion is evaluated for various values of $\breve{\vartheta}$ with $\breve{a} = 5$. In the first iteration, a point near $0.50$ is selected since the ${\rm IVAR}$ criterion is at its lowest. In the following iteration, points around $0.50$ become less favorable as they now yield higher ${\rm IVAR}$ values. The minimum shifts to around $0.92$, a point where the likelihood undergoes a significant change. Finally, a point from the high-level posterior region is chosen to improve the posterior prediction accuracy.

\subsection{Inference and Implementation}
\label{sec:inference}

At each stage, we obtain two distinct sets of $b$ parameters: one set is designed for exploitation purposes, as detailed in Section~\ref{sec:exploitation}, and the other to promote exploration, as outlined in Section~\ref{sec:exploration}. We then choose the strategy that results in the lowest overall posterior variance for evaluating the simulation model with $b$ parameters. 

Our derivations so far assume that intrinsic variance $r_j(\cdot)$ is known for all $j = 1, \ldots, \allowbreak d$. However, these values are not observed, and $r_j(\cdot)$ needs to be estimated at any parameter. Following \cite{Binois2018}, we use hetGP to model the log variances, denoted by $\log \mathbf{\Lambda}_{t,j}$, as the mean output of a GP on latent variables, $\mathbf{\Delta}_{t,j} = (\delta_1, \ldots, \delta_{n_t})^\top$. We use the Gaussian kernel, a widely used choice, in a separable form. The kernel function can be expressed as $c_{t,j}(\tg_i, \tg_{i'}) = \prod\limits_{\iota=1}^p \exp\left(-\frac{(\vartheta_{i,\iota} - \vartheta_{i',\iota})^2}{2\rho_{t,j,\iota}}\right)$, where the product across dimension $\iota$ enables independent scaling through the lenghtscales $\pmb{\rho}_{t,j} = (\rho_{t,j,1}, \ldots, \rho_{t,j,p})^\top$ in each dimension. The inverse of covariance matrix $\Kb_{t,j}(\ab_t)^{-1}$ can be defined as  $\Kb_{t,j}(\ab_t)^{-1} = \tau_{t,j}^{-1}(\Cb_{t,j} + \mathbf{A}_t^{-1}\mathbf{\Lambda}_{t,j})^{-1}$, with $[\Cb_{t,j}]_{ii'} = c_{t,j}(\tg_i, \tg_{i'})$ for $i, i' = 1, \ldots, n_t$, and $\mathbf{A}_t = {\rm diag}(a_1, \ldots, a_{n_t})$. The GP prior $\mathbf{\Delta}_{t,j} \sim {\rm MVN}\left(\mathbf{0}, \tau_{t,j}^g\left(\Cb_{t,j}^g + g_{t,j} \mathbf{A}_t^{-1}\right)\right)$ on the latent variables reveals that $\log \mathbf{\Lambda}_{t,j} = \Cb_{t,j}^g \left(\Cb_{t,j}^g + g_{t,j} \mathbf{A}_t^{-1}\right)^{-1} \mathbf{\Delta}_{t,j}$. Here, $\Cb_{t,j}^g$ represents the correlation matrix based on a Gaussian kernel function with lengthscale parameters $\pmb{\rho}^g_{t,j}$, $g_{t,j} > 0$ is the nugget parameter and $\tau_{t,j}^g$ is the scaling parameter. Inference involves choosing the values of $\pmb{\rho}_{t,j}$, $\pmb{\rho}^g_{t,j}$, $\mathbf{\Delta}_{t,j}$, and $g_{t,j}$. The joint log-likelihood is optimized by utilizing its gradient with respect to $\pmb{\rho}_{t,j}$, $\pmb{\rho}^g_{t,j}$, $\mathbf{\Delta}_{t,j}$, and $g_{t,j}$, as detailed in \cite{Binois2018}. Conditional on these values, the scaling parameters $\tau_{t,j}$ and $\tau_{t,j}^g$ have plug-in maximum likelihood estimates. Following its implementation in the R software package available on CRAN \citep{Binois2021}, we have developed a Python version within our package using the Gaussian kernel function. For the allocation rules presented in Sections~\ref{sec:exploitation}--\ref{sec:exploration}, $r_j(\tg_i)$ is estimated by $[\mathbf{\Lambda}_{t,j}]_{ii}$ if the intrinsic variance is needed for an existing point. Otherwise, the GP on latent variables is employed for estimation. We note that while the choice of kernel function does not affect the underlying rationale of our proposed acquisition functions, the Gaussian kernel’s strong smoothness assumptions may not always be appropriate for all simulation models. In such cases, alternative kernels, such as those from the Mat\'ern family, may better capture the variability of the underlying model. To enhance the flexibility of our code, we leave the implementation of additional kernel options as future work, which will allow users to select the most appropriate kernel for their specific use case.

\section{Results}
\label{sec:experiment}

Section~\ref{sec:synthetic} investigates the application of the IVAR criteria using three synthetic simulation models. In Section~\ref{sec:epidemic}, we demonstrate the implementation of the proposed approach with epidemiologic simulation models. We provide an open-source implementation of the batch sequential framework within the PUQ Python package at \hb{\url{https://github.com/parallelUQ/PUQ}}. All the test functions, along with the epidemiologic simulation models, can be found in the examples directory of the PUQ Python package.

\subsection{Results with synthetic functions}
\label{sec:synthetic}

We evaluate the performance of the proposed IVAR acquisition functions using three widely used synthetic simulation models: the unimodal, banana, and bimodal functions (see similar deterministic simulation models in \cite{Surer2023}). The parameter space for these models is two-dimensional ($p=2$), with the density patterns illustrated in Figures~\ref{fig:synth_IVAR_illus} and \ref{fig:synth_IVAR_illus_hybrid}. The simulation model generates one-dimensional output for the unimodal function ($d=1$) and two-dimensional output for the banana and bimodal functions ($d=2$). Detailed information about the data generation process for the test functions can be found in Supplementary Material~\ref{sec:expdesign}. In all examples in this section, we assume a uniform prior over the parameter ranges specified in Supplementary Material~\ref{sec:expdesign}. We expose each model to varying levels of noise intensity to evaluate the proposed approach across different signal-to-noise ratios. Figure~\ref{fig:synth_figs} displays the contour lines of intrinsic variance for each example. 

We evaluate the performance of our approach, which integrates exploration and exploitation strategies within the batch sequential procedure, using batch sizes $b \in \{8, 16, 32, 64\}$ to acquire 256 simulation runs.  Additional results focusing solely on the benefit of replication are provided in Supplementary Material~\ref{sec:experiment_replication}. For exploration, we generate a batch of $\breve{b}$ new parameters using Algorithm~\ref{alg:krigingbeliever} and allocate $\breve{a} = 2$ replicates, resulting in $b = \breve{b} \times \breve{a}$. As a benchmark, we derive ${\rm IMSE}(\ct, \breve{a})$ as detailed in Supplementary Material~\ref{proof:imse}. Subsequently, we replace ${\rm IVAR}(\ct, \breve{a})$ with ${\rm IMSE}(\ct, \breve{a})$ in Algorithm~\ref{alg:krigingbeliever} to generate a batch of $b$ parameters for exploration based on IMSE criterion. For replication, we use ${\rm IMSE}(\Delta \ab_t)$, as detailed in Supplementary Material~\ref{proof:imse_exploit}. Additionally, we draw inspiration from selecting inputs with the highest emulator variance to fit a global emulator \citep{Seo200}. Following this, at each iteration of Algorithm~\ref{alg:krigingbeliever}, we select parameters with the highest posterior uncertainty for exploration and allocate replicates based on our approach. We refer to this method as VAR in our benchmark. At each iteration of Algorithm~\ref{alg:krigingbeliever}, a candidate list of size $|\mathcal{L}_{\breve{t}}| = 200$ is generated randomly from the prior distribution. As part of our benchmark, we include a reference labeled UNIF, where we sample 64 new parameters using LHS and replicate each point four times. 

Our experimental setup uses an initial sample size of $n_0 = 15$ generated via LHS, with two replicates per point.  We follow the same setup described in Supplementary Material~\ref{sec:experiment_replication}, where for each experimental replicate, we rerandomize the observed data and the initial sample. The same observed data and initial sample are then used across all methods (IVAR, VAR, IMSE, and UNIF) within that replicate. Performance is evaluated using the mean absolute difference (MAD) between the estimated and true posteriors, as well as the interval score. Full definitions of these metrics are provided in Supplementary Material~\ref{sec:experiment_replication}.

\begin{figure}[ht]
\centering
    \begin{subfigure}{1\textwidth}    
        \includegraphics[width=1\textwidth]{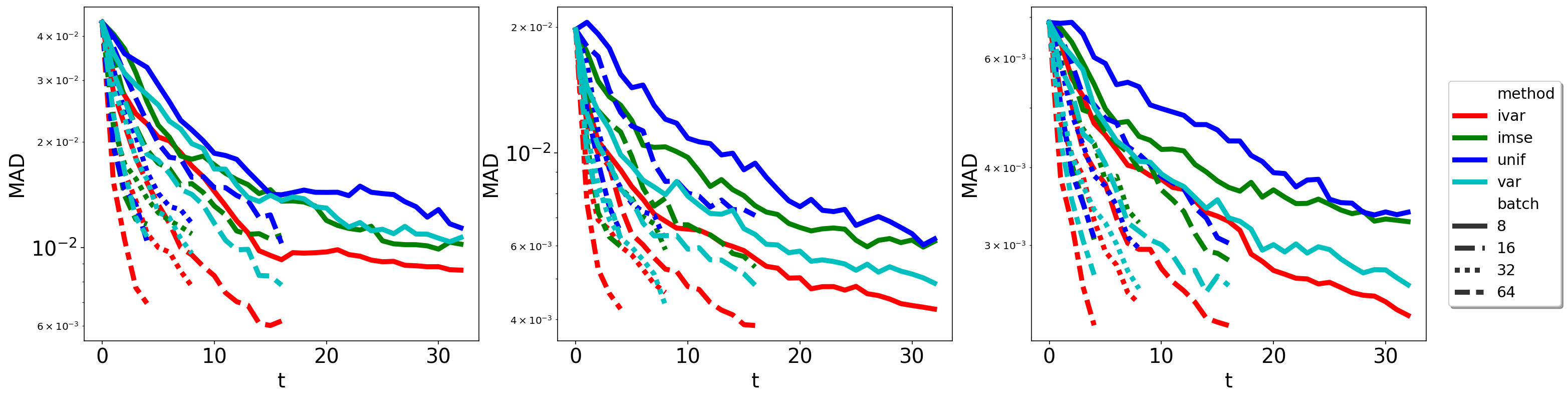}
    \end{subfigure}
    \caption{Comparison of different acquisition functions for unimodal (left), banana (middle), and bimodal (right) functions.}
    \label{fig:results_exps}
\end{figure}

\begin{figure}[t]
\centering
    \begin{subfigure}{1\textwidth}    
        \includegraphics[width=1\textwidth]{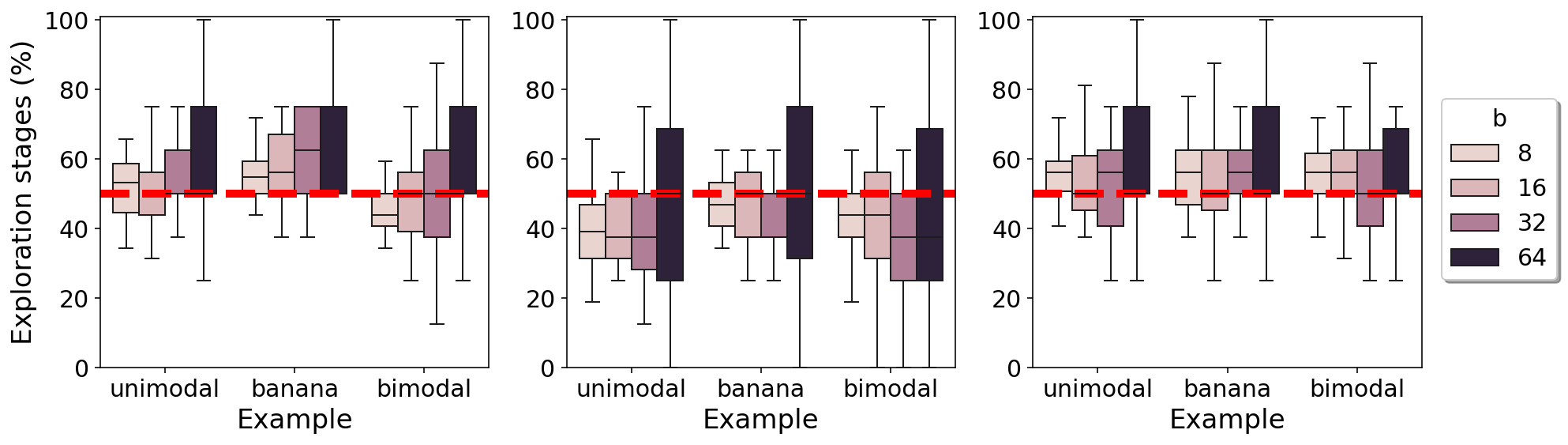}
    \end{subfigure}
    \caption{Percentage of exploration stages out of $T_b$ stages over 30 replications, displayed with IVAR (left), VAR (middle), and IMSE (right) acquisition functions.}
    \label{fig:percent}
\end{figure}
Figure~\ref{fig:results_exps} summarizes the MAD values across 30 experiment replicates.  In Supplementary Material~\ref{sec:additional_synt_res}, Figure~\ref{fig:results_exps_2} provides an alternative view of these results, and Table~\ref{tab:interval_score_explore} presents the corresponding interval scores over the same 30 replicates.
Figure~\ref{fig:synth_IVAR_illus_hybrid} illustrates the acquired parameters with $b=16$ for a single replicate of the experiments. 
IVAR prioritizes exploration within the calibration region of interest and allocates almost all replicates to this region. 
On the other hand, IMSE covers the entire parameter space to build a global emulator and encourages replicates from the region where the uncertainty is higher. 
As a result, the interval score is smaller with IVAR, ensuring that the allocated points are concentrated where they are most needed to learn the posterior better. Although VAR generally learns the posterior better than IMSE, it is more sensitive to intrinsic uncertainty compared to IVAR, as it focuses solely on the posterior variance at individual points. For instance, for the unimodal function in Figure~\ref{fig:synth_IVAR_illus_hybrid}, many points are drawn toward regions of high uncertainty, leaving the high-posterior region on the left-hand side underexplored. Furthermore, while VAR tends to concentrate parameters in the highest posterior regions (as in the bimodal function in Figure~\ref{fig:synth_IVAR_illus_hybrid}), IVAR selects parameters across a broader range—from low to high posterior regions—enhancing its ability to capture the posterior's overall shape.
The interval score of IVAR is generally larger compared to using only replicates, indicating that exploration identifies important regions not covered by the $10 \times 10$ grid used in the analysis in Supplementary Material~\ref{sec:experiment_replication}.
This indicates that space-filling designs can miss certain areas within the parameter region of interest, and combining exploration with exploitation helps better target these regions.
Moreover, integrating exploration with replication enhances the ability of the small batch size ($b=8$) to more effectively cover the region of interest, as evidenced by the reduced variability in interval scores. 
Figure~\ref{fig:percent} visualizes the percentage of stages dedicated to exploring the parameter space out of a total of $T_b$ stages to gain insights into the distribution of exploration and exploitation stages as batch sizes vary. 
In all examples, IVAR allocates the highest percentage of stages to exploration when using the largest batch size $b = 64$. 
However, having a higher percentage of stages dedicated to exploration does not necessarily lead to lower MAD values, as shown in Figure~\ref{fig:results_exps}. 
This is because IVAR encourages the selection of many new parameters over replicating existing ones with large batch sizes. 
In this work, we investigate a batch synchronous procedure where a set of parameters of size $b$ is obtained at each stage, and the process halts until all workers conclude $b$ simulation evaluations to acquire the next batch of $b$ parameters. 
Suppose one has access to many workers to evaluate the simulation model in parallel. 
In that case, the proposed acquisition functions can be integrated into a batch asynchronous procedure that does not require all simulations to finish before proceeding to the next stage. 
In such a case, the batch asynchronous procedure can take advantage of smaller batch sizes with a larger number of workers ($b < \text{the number of workers}$) to encourage more replicates.

We also assess the computational performance of the proposed acquisition functions, with details provided in Supplementary Material~\ref{sec:computational_performance}. Although acquisition costs are relatively low compared to simulation evaluations, efficient sampling techniques for integration over the parameter space and alternative emulators could further enhance computational efficiency.

\subsection{Application to epidemiological simulation models}
\label{sec:epidemic}

We illustrate the proposed sequential procedure with the discrete compartmental models in epidemiology used to simulate the spread of infectious diseases through populations. Compartmental models have been widely used during the COVID-19 pandemic to provide valuable insights into disease transmission dynamics and assess the potential impact of interventions such as social distancing, mask-wearing, and vaccination campaigns \citep{Yang2020}. The Susceptible-Infected-Recovered (SIR) model represents the most basic form of such a model. In this model, the infection rate governs the flow of individuals from the susceptible (S) compartment to the infected (I) compartment, while the removal rate determines the transition of individuals from the infected (I) compartment to the recovered (R) compartment. Given specific values for the infection and removal rate, the SIR model returns the susceptible, infected, and recovered individuals over time. In this work, in addition to the SIR model, we demonstrate the proposed approach with the enhanced Susceptible-Exposed-Infected-Recovered-Death-Susceptible (SEIRDS) model to provide an example in higher dimensional input and output spaces. Following the implementation presented in the R package by \cite{FitzJohn2023}, we implement both the SIR and SEIRDS models in our Python package. 

\begin{figure}[t]
\centering
    \begin{subfigure}{0.8\textwidth}    
        \includegraphics[width=1\textwidth]{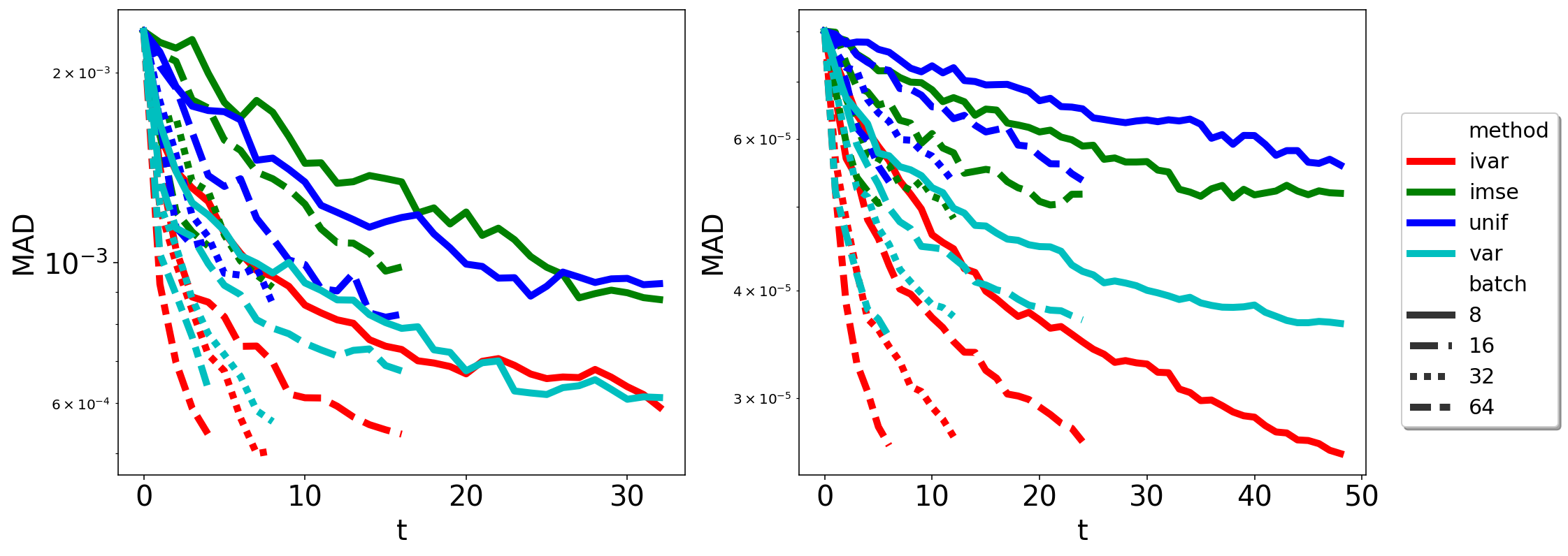}
    \end{subfigure}
    \caption{Comparison of different acquisition functions for SIR (left) and SEIRDS (right) models.}
    \label{fig:results_SIR}
\end{figure}
SIR and SEIRDS models have two- and seven-dimensional parameters (i.e., $p=2$ and $p=7$, respectively) controlling the transition across compartments. We assume a uniform prior for each parameter.  In Supplementary Material~\ref{sec:epi_illustrate}, Table~\ref{tab:sir_params} summarizes the parameters for both models. Each parameter is scaled to $[0, 1]$ to simplify the integrals. In the SIR model, we set the initial number of susceptible individuals to 1000 and the initial number of infected individuals to 10, within a population of 1010 individuals. The simulation runs for 150 days. Likewise, for the SEIRDS model, we begin with 1000 susceptible individuals and 10 exposed individuals and run the model for 150 days. We consider the average number of individuals within each compartment over the simulation period as the output. Consequently, the SIR and SEIRDS models yield outputs of three and six dimensions (i.e., $d=3$ and $d=6$, respectively). We model the observed data $\yb = \latent(\tb=\tb^*) + \epsilonv$ with $\epsilonv \sim {\rm MVN}(\mathbf{0}, \Sigmav)$. We set $\tb^*$ to the midpoint of the prior ranges, which are default values provided in the R package by \cite{FitzJohn2023}. Since the expected simulation output $\latent(\tb=\tb^*)$ is unknown, we substitute it with the average simulation output across thousands of replicates.

The sequential procedure begins by drawing an initial sample from LHS with a size of $n_0 = 15$ for the SIR model and $n_0 = 50$ for the SEIRDS model. 
Each point in the initial sample is replicated twice. 
We compare IVAR, VAR, and IMSE by selecting either exploration or replication at each stage. 
We obtain 256 and 384 parameters using each acquisition function for the SIR and SEIRDS models, respectively. 
As a baseline, we sample 64 and 96 parameters from LHS for the SIR and SEIRDS models, each with four replicates, labeled as UNIF in the benchmark. 
Each parameter acquired via Algorithm~\ref{alg:krigingbeliever} is selected from a candidate list of size $|\mathcal{L}_{\breve{t}}| = 200$ for SIR models and $|\mathcal{L}_{\breve{t}}| = 500$ for SEIRDS models, using $\breve{a}=2$. 
We generate a $50 \times 50$ grid of two-dimensional parameters for the SIR model as a reference set. 
For each parameter in the reference set, we replicate the simulation model a thousand times. 
The average simulation output across replicates is then employed as the expected simulation output to compute the true posterior $\tilde{p}(\thetav|\yb)$. 
Because a grid in seven-dimensional space is large, we utilize a sample of 2500 parameters from LHS as the reference set for the SEIRDS model. 
On each replicate of the experiments, we regenerate the observed data $\yb$ and the initial sample. 
We then acquire parameters using IVAR, VAR, IMSE, and UNIF all based on the same observed data $\yb$ and initial sample. 
We summarize the performance of the functions over the 30 replicates of the experiments by averaging $\text{MAD}_t$ values.
\begin{figure}[h]
\centering
    \begin{subfigure}{1\textwidth}    
        \includegraphics[width=1\textwidth]{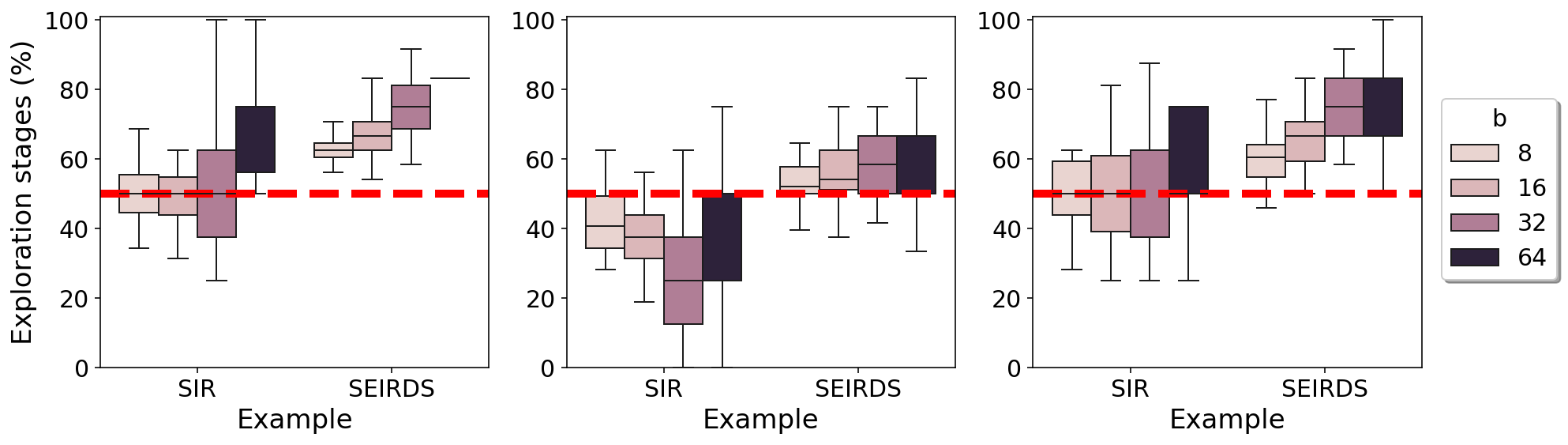}
    \end{subfigure}
    \caption{Percentage of exploration stages out of $T_b$ stages over 30 replications, displayed with IVAR (left), VAR (middle), and IMSE (right) acquisition functions.}
    \label{fig:percent_SIR}
\end{figure}

Figure~\ref{fig:results_SIR} presents the posterior prediction accuracy across various batch sizes.  Figure~\ref{fig:results_SIR_2} in Supplementary Material~\ref{sec:epi_illustrate} offers an alternative perspective on these results.
Additionally, Figure~\ref{fig:percent_SIR} illustrates the percentage of exploration stages for each batch size. 
Consistent with our findings in Section~\ref{sec:synthetic}, IVAR tends to explore more than replicate when the batch size is $b=64$. 
Additionally, the SEIRDS model generally benefits more from exploration than replication.
In Supplementary Material~\ref{sec:epi_illustrate}, Figure~\ref{fig:params_SIR} displays the parameters obtained using IVAR, VAR, and IMSE for a single replicate of the SIR experiments. Figures~\ref{fig:outputs_SIR}--\ref{fig:outputs_SEIRDS} in the same section show the simulation outputs for a single replicate of the IVAR, VAR, and IMSE experiments, averaged over 1000 replications. Both IVAR and VAR gather outputs concentrated around the expected simulation output at $\tb^*$. In contrast, IMSE selects many outputs from outside the region of interest. While VAR densely gathers outputs near the expected simulation output, IVAR distributes the outputs more broadly, since the acquired points span both low and high posterior regions. Consequently, IVAR facilitates more accurate posterior learning by prioritizing alignment with the observed data and distributing parameters across both high- and low-posterior regions. The relatively smaller interval scores of IVAR,  shown in Table~\ref{tab:interval_score_SIR} in Supplementary Material~\ref{sec:epi_illustrate}, compared to IMSE, provide further evidence of IVAR's ability to precisely target the region of interest.
Meanwhile, the relatively higher interval scores of IVAR, compared to VAR, highlight IVAR's strength in achieving broader and more comprehensive coverage within the region of interest.
An exception occurs for the SIR model when $b=8$, where IVAR and VAR perform similarly based on MAD values in Figure~\ref{fig:results_SIR}. In this case, VAR makes smaller errors in regions with higher posterior values due to its tendency to overexploit those areas. However, despite failing to effectively learn regions with lower posterior values, its average performance remains comparable to that of IVAR.

\section{Conclusion}
\label{sec:conc}

We propose a batch sequential approach for high-dimensional stochastic simulation models for improved posterior learning. The proposed acquisition functions allow a batch of simulation runs to be allocated to either explore the parameter space or replicate the existing parameters in the simulation data. Unlike existing approaches that primarily rely on intrinsic uncertainties to construct a global emulator, our criteria consider both the posterior density and uncertainties to better infer the parameters by minimizing the total posterior uncertainty.
Further avenues for research are anticipated. 
In parallel with how \cite{Surer2024} generalizes the integrated variance criterion introduced in \cite{Surer2023} for high-dimensional deterministic outputs, the proposed approach can be extended to one-dimensional simulation outputs to acquire both design points and parameters.
While batch updates are generally considered more computationally efficient than one-at-a-time sequential updates, the latter often yield more accurate results due to their ability to make more informed decisions at each stage based on the most recent information. Further investigation into the accuracy of one-at-a-time updates compared to proposed batch updates remains a promising avenue.  Another interesting direction for future exploration is the integration of exploration and replication within the same batch, rather than fully separating them into distinct sets of parameters. An alternative approach could involve first determining the total number of simulation runs (the ``budget'') and then distributing this budget between exploration and exploitation. Further research into this approach, including the development of decision rules for effective resource allocation within a batch, could provide valuable insights into enhancing optimization efficiency and effectiveness in batch sequential frameworks. Moreover, we assume user-defined values for the number of new design points, $\breve{b}$, and replicates, $\breve{a}$, when exploring the parameter space. Suitable recommendations for the selection of $\breve{a}$ and $\breve{b}$ are open for study. 
In this work, we assume that collecting new observed data is not possible, and therefore, the existing observed data remains unchanged throughout the procedure. Following this work, we further explore extensions for physical experiment design and the integration of simulation and physical experiments. Deciding on replication and exploration in the design of physical experiments is crucial for more accurate and robust inference and better resource utilization. 
Finally, several critical aspects of batch sequential design impact performance assessment, such as batch size, simulation model runtime, convergence criteria, and acquisition function runtime. A batch size that excels in one application may perform poorly in another, highlighting the need to tailor strategies to specific contexts. Analytic performance models, such as those proposed by \cite{Surer2025,surer2024active}, can provide valuable insights into the trade-offs between candidate configurations, helping practitioners select the most suitable approach for their unique problems and settings.

\appendix

\section*{Supplementary Material}

\counterwithin{figure}{section}
\counterwithin{equation}{section}
\renewcommand\thefigure{\thesection.\arabic{figure}} 
\renewcommand\theequation{\thesection.\arabic{equation}} 

\input{appABCD}

\input{appE}

\input{appF}

\input{appG}

\input{appH}

\input{appI}

\input{appJ}

\section*{Acknowledgement}
\hb{The author is grateful to Matthew T. Pratola for the valuable discussion. The author gratefully acknowledges the computing resources provided on Bebop, a high-performance computing cluster operated by the Laboratory Computing Resource Center at Argonne National Laboratory.}

\subsection*{Funding}
\hb{\"{O}zge S\"{u}rer is grateful for support from the National Science Foundation (NSF) grant OAC 2004601.}

\clearpage
\bibliographystyle{plainnat}
\bibliography{activelearning}

\end{document}

%% file: macros.tex
\usepackage{caption}
\usepackage{subcaption}
\usepackage{graphicx}

\usepackage{times,scalefnt}
\usepackage{amsmath,amssymb,mathtools}
\usepackage{tikz}
\usetikzlibrary{matrix,decorations.pathreplacing, calc, positioning, shapes.callouts, shapes.geometric}

\usepackage{amsmath}

\DeclareMathOperator*{\argmin}{arg\,min}

\usepackage[boxruled,vlined,linesnumbered,commentsnumbered]{algorithm2e}

\usepackage{subcaption}

\newcommand{\p}{{p}}

\newcommand{\Cb}{{\pmb{C}}}

\newcommand{\kb}{{\mathbf{k}}} 
\newcommand{\Kb}{{\mathbf{K}}} 

\newcommand{\y}{{y}}
\newcommand{\yb}{{\mathbf{y}}}

\newcommand{\simout}{{\pmb{\zeta}}}
\newcommand{\latent}{{\pmb{\eta}}}

\newcommand{\epsilonv}{{\pmb{\varepsilon}}}
\newcommand{\thetav}{{\pmb{\theta}}}
\newcommand{\Sigmav}{{\pmb{\Sigma}}}

\newcommand{\Vb}{{\mathbf{V}}}

\newcommand{\Rb}{{\mathbf{R}}}
\newcommand{\Sv}{{\mathbf{S}}}

\newcommand{\Mb}{{\mathbf{M}}}

\newcommand{\xb}{{\mathbf{x}}}

\newcommand{\tb}{{\pmb{\theta}}}
\newcommand{\ct}{{\breve{\pmb{\vartheta}}}}
\newcommand{\tg}{{\pmb{\vartheta}}}
\newcommand{\PHI}{{\pmb{\phi}}}

\newcommand{\muv}{{\pmb{\mu}}}
\newcommand{\nuv}{{\pmb{\nu}}}

\newcommand{\ab}{{\pmb{a}}}

\newcommand{\Exp}{\mathbb{E}_{\bar{\breve{\simout}} | \mathcal{D}_t}}

\newcommand{\DatExt}{\breve{\mathcal{D}}_{t}}
\newcommand{\DatExtR}{\mathcal{D}^\Delta_{t}}

\newcommand{\E}{\mathbb{E}_{\latent(\tb) | \mathcal{D}_{t}}}
\newcommand{\V}{\mathbb{V}_{\latent(\tb) | \mathcal{D}_{t}}}
\newcommand{\VE}{\mathbb{V}_{{\latent(\tb) | \DatExt}}}
\newcommand{\VR}{\mathbb{V}_{{\latent(\tb) | \DatExtR}}}

%% file: appABCD.tex
\section{Derivation of IMSE Criterion for Replication}
\label{proof:imse_exploit}
The integrated mean squared error ${\rm IMSE}(\Delta \ab_t)$ minimizes $\sum\limits_{\tb \in \Theta_{\rm ref}}  \sum\limits_{j=1}^d \left[\mathbf{S}_t^\Delta(\tb)\right]_{j,j}$ as a function of the number of additional replicates allocated to each design point. One can replace the objective function ${\rm IVAR}(\Delta \ab_t)$ in \eqref{prob:repopt} with the ${\rm IMSE}(\Delta \ab_t)$ to find the optimal allocation and obtain
\begin{align} \notag
    \begin{split}
        a_i + \Delta a_i \propto \sqrt{C_{i}^{\rm IMSE}(\ab_t)}, \quad i = 1, \ldots, n_t,
    \end{split}
\end{align}
where $C_{i}^{\rm IMSE}(\ab_t) = \sum\limits_{\tb \in \Theta_{\rm ref}} \sum\limits_{j=1}^d  \left[\Mb_i(\tb)\right]_{j,j}$. Recall that $\Mb_i(\tb)$ is a $d \times d$ diagonal matrix with $j$th diagonal element $-r_j(\tg_i)\kb^\top_{t,j}(\tb) \Kb_{t,j}(\ab_t)^{-1}  \mathbf{J}^{(i,i)} \Kb_{t,j}(\ab_t)^{-1} \kb_{t,j}(\tb)$. Note that this criterion is equivalent to the criterion introduced by \cite{Chen2017} approximated over a discrete set of parameters for $j=1$. 

\section{Proof of Theorem~3.2}
\label{proof:ivar_rep}

We first formulate the Lagrangian $L(\Delta \ab_t)$ as 
\begin{align} \notag 
    \begin{split}
        L(\Delta \ab_t) = {\rm IVAR}(\Delta \ab_t) + \lambda \left(b - \sum_{i=1}^{n_t} \Delta a_i\right).
    \end{split}
\end{align}  

The first-order optimality conditions are 
\begin{align} \notag 
    \begin{split}
        \frac{\partial L(\Delta \ab_t)}{\partial \Delta a_i} = \frac{\partial  {\rm IVAR}(\Delta \ab_t)}{\partial \Delta a_i} -\lambda = 0, \quad i = 1, \ldots, n_t.
    \end{split}
\end{align}     
We have
\begin{align} \notag 
    \begin{split}
        \frac{\partial {\rm IVAR}(\Delta \ab_t)}{\partial \Delta a_i} = \frac{1}{2^d\pi^{d/2}|\Sigmav|^{1/2}} \int\limits_{\tb \in \Theta} \frac{\partial f(\tb)}{\partial \Delta a_i} p(\tb)^2 d\tb - 2 \int\limits_{\tb \in \Theta} g(\tb) \frac{\partial g(\tb)}{\partial \Delta a_i} p(\tb)^2 d\tb,
    \end{split}
\end{align}
where 
\[
    \frac{\partial f(\tb)}{\partial \Delta a_i} = f(\tb) \frac{\partial \log f(\tb)}{\partial \Delta a_i} \quad {\rm and} \quad \frac{\partial g(\tb)}{\partial \Delta a_i} = g(\tb) \frac{\partial \log g(\tb)}{\partial \Delta a_i}.
\]
Define $\dot{\mathbf{N}}(\tb) = 0.5\Sigmav + \Sv_{t}^\Delta(\tb)$, $\mathbf{N}(\tb) = \Sigmav + \Sv_{t}^\Delta(\tb)$, and $\mathbf{h}(\tb) = \yb-\muv_{t}(\tb)$. By taking partial derivatives, we can write
\begin{align} \label{partialfandg}
    \begin{split}
    &\frac{\partial \log f(\tb)}{\partial  \Delta  a_i} = -\frac{1}{2} \frac{\partial \log |\dot{\mathbf{N}}(\tb)|}{\partial \Delta a_i} -\frac{1}{2}\mathbf{h}(\tb)^\top\frac{\partial\dot{\mathbf{N}}(\tb)^{-1}}{\partial \Delta a_i}\mathbf{h}(\tb), \\
    &\frac{\partial \log g(\tb)}{\partial \Delta a_i} = -\frac{1}{2} \frac{\partial \log |\mathbf{N}(\tb)|}{\partial \Delta a_i} -\frac{1}{2}\mathbf{h}(\tb)^\top\frac{\partial \mathbf{N}(\tb)^{-1}}{\partial \Delta a_i}\mathbf{h}(\tb).
    \end{split}
\end{align}
From matrix calculus, we have $\frac{\partial \log |\mathbf{Y}|}{\partial x} = {\rm Tr}\left(\mathbf{Y}^{-1} \frac{\partial \mathbf{Y}}{\partial x}\right)$ and $\frac{\partial \mathbf{Y}^{-1}}{\partial x} = -\mathbf{Y}^{-1} \frac{\partial \mathbf{Y}}{\partial x} \mathbf{Y}^{-1}$. Then, \eqref{partialfandg} is equivalently written
\begin{align} \label{eq:logderivative}
    \begin{split}
         \frac{\partial \log f(\tb)}{\partial \Delta a_i} &= -\frac{1}{2} {\rm Tr}\left(\dot{\mathbf{N}}(\tb)^{-1}\frac{\partial \dot{\mathbf{N}}(\tb)}{\partial \Delta a_i} \right) + \frac{1}{2}\mathbf{h}(\tb)^\top \dot{\mathbf{N}}(\tb)^{-1} \frac{\partial \dot{\mathbf{N}}(\tb)}{\partial \Delta a_i}\dot{\mathbf{N}}(\tb)^{-1} \mathbf{h}(\tb), \\
         \frac{\partial \log g(\tb)}{\partial \Delta a_i} &= -\frac{1}{2} {\rm Tr}\left(\mathbf{N}(\tb)^{-1}\frac{\partial \mathbf{N}(\tb)}{\partial \Delta a_i} \right) + \frac{1}{2}\mathbf{h}(\tb)^\top \mathbf{N}(\tb)^{-1} \frac{\partial \mathbf{N}(\tb)}{\partial \Delta a_i}\mathbf{N}(\tb)^{-1} \mathbf{h}(\tb),
    \end{split}
\end{align}
where $\frac{\partial \dot{\mathbf{N}}(\tb)}{\partial \Delta a_i} = \frac{\partial \mathbf{N}(\tb)}{\partial \Delta a_i} = \frac{\partial \Sv_{t}^\Delta(\tb)}{\partial \Delta a_i} = {\rm diag}\left(\frac{\partial \varsigma^2_{t,1}(\tb, \Delta \ab_t)}{\partial \Delta a_i}, \ldots, \frac{\partial \varsigma^2_{t,d}(\tb, \Delta \ab_t)}{\partial \Delta a_i}\right)$. Here, $\varsigma^2_{t,j}(\tb, \Delta \ab_t)$ is the emulator variance in \eqref{eq:meanvar_latent} obtained with additional $\Delta \ab_t$ replicates.

Define $\Vb_{j}(\Delta\ab_t) = {\rm{diag}}\left(\frac{r_j(\tg_1)}{a_1 + \Delta a_1}, \ldots,  \frac{r_j(\tg_{n_t})}{a_{n_t} + \Delta a_{n_t}}\right)$ such that $\Vb_{j}(\Delta\ab_t) = \Vb_{j}(\ab_t) + \Delta\Vb_j(\ab_t)$ where $\Vb_j(\ab_t) = {\rm{diag}}\left(\frac{r_j(\tg_1)}{a_1}, \ldots,  \frac{r_j(\tg_{n_t})}{a_{n_t}}\right)$ and $\Delta\Vb_j(\ab_t) = {\rm{diag}}\left(\frac{-r_j(\tg_1)\Delta a_1}{(a_1 + \Delta a_1)a_1}, \ldots, \frac{-r_j(\tg_{n_t})\Delta a_{n_t}}{(a_{n_t} + \Delta a_{n_t})a_{n_t}}\right)$. Denote the covariance matrix $\Kb_{t,j}(\Delta \ab_t) = \Kb_{t,j} + \Vb_j(\Delta \ab_t) = \Kb_{t,j}(\ab_t) + \Delta\Vb_j(\ab_t)$ after adding $\Delta \ab_t$ replications. The Woodbury matrix identity gives 
\begin{align} 
    \begin{split} \label{eq:varderivative1}
    \frac{\partial \varsigma^2_{t,j}(\tb, \Delta \ab_t)}{\partial \Delta a_i} &= - \kb^\top_{t,j}(\tb) \frac{\partial \Kb_{t,j}(\Delta \ab_t)^{-1}}{\partial \Delta a_i} \kb_{t,j}(\tb) \quad \rm{where} \\
    \Kb_{t,j}(\Delta \ab_t)^{-1} &= \Kb_{t,j}(\ab_t)^{-1} - \Kb_{t,j}(\ab_t)^{-1} \left( \mathbf{I} + \Delta \Vb_j(\ab_t) \Kb_{t,j}(\ab_t)^{-1} \right)^{-1} \Delta \Vb_j(\ab_t) \Kb_{t,j}(\ab_t)^{-1}.
    \end{split}
\end{align}
For brevity, let $\mathbf{A} = \Delta \Vb_j(\ab_t) \Kb_{t,j}(\ab_t)^{-1}$. Finally, the application of matrix calculus gives 
\begin{align} \label{eq:derivative}
    \begin{split} 
   & \frac{\partial}{\partial \Delta a_i} \left(\Kb_{t,j}(\ab_t)^{-1} \left(\mathbf{I} + \mathbf{A} \right)^{-1} \mathbf{A}\right)  \\
   & = \Kb_{t,j}(\ab_t)^{-1} \left( \mathbf{I} + \mathbf{A}\right)^{-1} \frac{r_j(\tg_i)}{(a_i + \Delta a_i)^2} \mathbf{J}^{(i,i)} \Kb_{t,j}(\ab_t)^{-1} \left( \mathbf{I} + \mathbf{A} \right)^{-1}\mathbf{A} \\
   & -\Kb_{t,j}(\ab_t)^{-1} \left( \mathbf{I} + \mathbf{A} \right)^{-1} \frac{r_j(\tg_i)}{(a_i + \Delta a_i)^2} \mathbf{J}^{(i,i)} \Kb_{t,j}(\ab_t)^{-1}.
    \end{split}
\end{align}
Similar to the assumption presented in Theorem~1 of \cite{Chen2017}, we also assume that $\max \limits_{i=1, \ldots, n_t} \left[\Delta \mathbf{V}_j(\ab_t)\right]_{i,i} \ll 1$ such that $\mathbf{I} + \Delta \Vb_j(\ab_t) \Kb_{t,j} (\ab_t)^{-1} \approx \mathbf{I}$. Then, \eqref{eq:derivative} can be further simplified as
\begin{align} \label{eq:simplederivative}
    \begin{split} 
        & \frac{\partial}{\partial \Delta a_i} \left(\Kb_{t,j}(\ab_t)^{-1} \left(\mathbf{I} + \mathbf{A} \right)^{-1} \mathbf{A}\right) \approx -\frac{r_j(\tg_i)}{(a_i + \Delta a_i)^2}\Kb_{t,j}(\ab_t)^{-1}  \mathbf{J}^{(i,i)} \Kb_{t,j}(\ab_t)^{-1}.
    \end{split}
\end{align}

Substituting \eqref{eq:simplederivative} into \eqref{eq:varderivative1} gives
\begin{align} \label{eq:varderivative2}
    \begin{split} 
        \frac{\partial \varsigma^2_{t,j}(\tb, \Delta \ab_t)}{\partial \Delta a_i} = -\frac{r_j(\tg_i)}{(a_i + \Delta a_i)^2}\kb^\top_{t,j}(\tb) \Kb_{t,j}(\ab_t)^{-1}  \mathbf{J}^{(i,i)} \Kb_{t,j}(\ab_t)^{-1} \kb_{t,j}(\tb).
    \end{split}
\end{align}
Moreover, in \eqref{eq:logderivative}, $\dot{\mathbf{N}}(\tb) \approx 0.5\Sigmav + \Sv_{t}(\tb)$, $\mathbf{N}(\tb) \approx \Sigmav + \Sv_{t}(\tb)$ since $\Sv_{t}^\Delta(\tb)$ can now be approximated with $\Sv_{t}(\tb)$ due to our assumption.
Finally, plugging \eqref{eq:varderivative2} into \eqref{eq:logderivative} and then solving first-order optimality conditions completes the proof.

\section{Derivation of IMSE Criterion for Exploration}
\label{proof:imse}

As a benchmark in the experiments, we replace ${\rm IVAR}(\ct, \breve{a})$ in Algorithm~\ref{alg:krigingbeliever} with ${\rm IMSE}(\ct, \breve{a})$, where ${\rm IMSE}(\ct, \breve{a})$ is defined as $\sum\limits_{\tb \in \Theta_{\rm ref}} \sum\limits_{j=1}^d \left[\mathbf{S}_{t+1}(\tb)\right]_{j,j}$. Here, $\mathbf{S}_{t+1}(\cdot)$ represents the covariance matrix after observing unknown simulation data points $\left(\left(\ct, \simout(\ct)^l\right) : l = 1, \ldots, \breve{a}\right)$.
Recall that $m_{t,j}(\cdot)$ and $\varsigma^2_{t,j}(\cdot)$ in \eqref{eq:meanvar_latent} refer to the emulator mean and variance at stage $t$ for output $j = 1, \ldots, d$. After observing the hypothetical data points, we can write
\begin{align}\label{updatedK}
    \begin{split}
        \Kb_{t+1,j}(\ab_{t+1}) = \left[ {\begin{array}{cc} \Kb_{t,j}(\ab_t) & \kb_{t,j}(\ct) \\
                        \kb_{t,j}(\ct)^\top & k_{t,j}(\ct, \ct) + \frac{r_j\left(\ct\right)}{\breve{a}} \\
                        \end{array} } \right],
    \end{split}      
\end{align}
where $\ab_{t+1} = (a_1, \ldots, a_{n_t}, \breve{a})^\top$.
The partition inverse equations give the matrix inverse $\Kb_{t+1,j}(\ab_{t+1})^{-1}$ as
\begin{align}\notag
    \begin{split}
         \breve{l} \left[{\begin{array}{cc} \Kb_{t,j}(\ab_t)^{-1} \breve{l}^{-1} + \Kb_{t,j}(\ab_t)^{-1} \kb_{t,j}(\ct) \kb_{t,j}(\ct)^\top \Kb_{t,j}(\ab_t)^{-1} & -\Kb_{t,j}(\ab_t)^{-1} \kb_{t,j}(\ct) \\
            -\kb_{t,j}(\ct)^\top \Kb_{t,j}(\ab_t)^{-1} & 1\\
            \end{array} } \right]
    \end{split}      
\end{align}
where $\breve{l} = \left(\varsigma^2_{t,j}(\ct) + \frac{r_j(\ct)}{\breve{a}}\right)^{-1} $. Let $\varsigma^2_{t+1,j}(\tb)$ be the variance at $\tb$ after seeing $\left(\ct, \simout(\ct)^l\right)$, for $l = 1, \ldots, \breve{a}$. Then, the variance function yields
    \begin{align}\label{var_future}
        \begin{split}
            \varsigma^2_{t+1,j}(\tb) &= k_{t,j}(\tb, \tb) - \left[\kb_{t,j}^\top(\tb), k_{t,j}(\tb, \ct)\right]^\top \Kb_{t+1,j}(\ab_{t+1})^{-1} \left[ {\begin{array}{cc} \kb_{t,j}(\tb)\\
            k_{t,j}(\tb, \ct) \\
                \end{array} } \right] \\
            &= \varsigma^2_{t,j}(\tb) - \breve{l} {\rm cov}_{t,j}(\tb, \ct)^2.
        \end{split} 
    \end{align}
Let $\PHI_t(\tb, \ct)$ be the $d \times d$ diagonal matrix with diagonal elements $\frac{\text{cov}_{t,j}(\tb, \ct)^2}{\varsigma^2_{t,j}(\ct) + r_j(\ct)/\breve{a}}$. Then, we have $\mathbf{S}_{t+1}(\tb) = \mathbf{S}_{t}(\tb) - \PHI_t(\tb, \ct)$. Thus, we compute ${\rm IMSE}(\ct, \breve{a})$ for each $\ct \in \mathcal{L}_{\breve{t}}$ and choose the one that minimizes the IMSE to include in the batch.

\section{Proof of Lemma~3.3}
\label{proof:lemma3.3}


In Section~\ref{proof:imse}, we derive the variance of the emulator $\varsigma^2_{t+1,j}(\tb)$ after adding a new parameter $\ct$ with (unseen) $\breve{a}$ replicates to the simulation data. To compute the criterion, we continue to establish general results for GPs. Recall that $\bar{\breve{\simout}} = \left(\bar{\breve{\zeta}}_1, \ldots, \bar{\breve{\zeta}}_d\right)^\top $ represents the unseen sample average simulation output at a new parameter $\ct$, where $\bar{\breve{\zeta}}_j = \frac{\sum\limits_{l=1}^{\breve{a}}\zeta(\xb_j, \ct)^l}{\breve{a}}$ for $j = 1, \ldots, d$. Utilizing a similar approach in \eqref{var_future}, we begin by determining the mean of the emulator $m_{t+1,j}(\tb)$ after incorporating a new parameter $\ct$ with (unseen) $\breve{a}$ replicates into the simulation data such that
\begin{align}\label{mean_future}
    \begin{split}
        m_{t+1,j}(\tb) &= \left[\kb_{t,j}^\top(\tb), k_{t,j}(\tb, \ct)\right] \Kb_{t+1,j}(\ab_{t+1})^{-1}  \left[ {\begin{array}{cc} \bar{\simout}_{t,j} \\
        \bar{\breve{\zeta}}_j \\
            \end{array} } \right] \\
        &= m_{t,j}(\tb) + \breve{l} \text{cov}_{t,j}(\tb, \ct)\left(\bar{\breve{\zeta}}_j - m_{t,j}(\ct)\right),
    \end{split} 
\end{align}
where $\breve{l} = \left(\varsigma_{t,j}^2(\ct) + \frac{r_j(\ct)}{\breve{a}}\right)^{-1}$.
Taking the expected value and the variance of \eqref{mean_future}, respectively, provides
    \begin{align}
        \begin{split}
        \mathbb{E}_{\bar{\breve{\zeta}}_j|\mathcal{D}_t}\left[m_{t+1,j}(\tb)\right] = m_{t,j}(\tb), \quad 
        \mathbb{V}_{\bar{\breve{\zeta}}_j|\mathcal{D}_t}\left[m_{t+1,j}(\tb)\right] = \frac{\text{cov}_{t,j}(\tb, \ct)^2}{\varsigma^2_{t,j}(\ct) + r_j(\ct)/\breve{a}}. \\
    \end{split}
\end{align}
Using $\bar{\breve{\zeta}}_j|\mathcal{D}_t \sim \text{N}\left(m_{t,j}(\ct), \, \varsigma^2_{t,j}(\ct) + r_j(\ct)/\breve{a}\right)$ and the transformation in \eqref{mean_future}, we have
    \begin{align}\label{mean_distr1}
        \begin{split}
            m_{t+1,j}(\tb) | \mathcal{D}_t \sim \text{N}\left(m_{t,j}(\tb), \, \frac{\text{cov}_{t,j}(\tb, \ct)^2}{\varsigma^2_{t,j}(\ct) + r_j(\ct)/\breve{a}}\right).
       \end{split} 
    \end{align}
Recall that $\PHI_t(\tb, \ct)$ is the $d \times d$ diagonal matrix with diagonal elements $\frac{\text{cov}_{t,j}(\tb, \ct)^2}{\varsigma^2_{t,j}(\ct) + r_j(\ct)/\breve{a}}$ and $\muv_{t}(\tb)$ represents the mean vector of simulation outputs at stage $t$. Then, \eqref{mean_distr1} implies 
    \begin{align}\label{mean_distr2}
        \begin{split}
            \muv_{t+1}\left(\tb\right) | \mathcal{D}_t \sim \text{MVN}\left(\muv_{t}(\tb), \PHI_t(\tb, \ct)\right).
       \end{split} 
    \end{align}  

To derive ${\rm IVAR}(\ct, \breve{a})$, we first compute the expectation $\Exp\left(\VE\left[\tilde{p}(\tb|\yb)\right]\right)$, where the extended data set $\DatExt$ is given by $\DatExt = \mathcal{D}_t $ $\cup \left(\left(\ct, \simout(\ct)^l\right) : l = 1, \ldots, \breve{a}\right)$.        
We calculate the variance $\VE\left[\tilde{p}(\tb|\yb)\right]$ as
\[
    \VE\left[\tilde{p}(\tb|\yb)\right] = \VE\left[\tilde{p}(\yb|\tb)p(\tb)\right] = \VE\left[\tilde{p}(\yb|\tb)\right]p(\tb)^2,
\] where, due to Lemma~\ref{lemma:UQ}, $\VE\left[\tilde{p}(\yb|\tb)\right]$ is given by
    \begin{align} \notag 
        \begin{split}
             & \frac{1}{2^{d}\pi^{d/2}|\Sigmav|^{1/2}}f_\mathcal{N}\left(\yb; \, \muv_{t+1}(\tb), \, \frac{1}{2}\Sigmav + \Sv_{t+1}(\tb)\right)  - \left(f_\mathcal{N}\left(\yb; \, \muv_{t+1}(\tb), \, \Sigmav + \Sv_{t+1}(\tb)\right)\right)^2.
        \end{split}
    \end{align}

Due to \eqref{var_future}, we have $\mathbf{S}_{t+1}(\tb) = \mathbf{S}_{t}(\tb) - \PHI_t(\tb, \ct)$ and $\mathbf{S}_{t+1}(\tb)$ does not depend on $\bar{\breve{\simout}}$. By replacing $\mathbf{S}_{t+1}(\tb)$ with $\mathbf{S}_{t}(\tb) - \PHI_t(\tb, \ct)$ and using \eqref{mean_distr2}, we obtain $\Exp\left(\VE\left[\tilde{p}(\yb|\tb)\right]\right)$ 
    \begin{align} \label{eq:IVAR_explore}
        \begin{split}
             & \int \frac{1}{2^{d}\pi^{d/2}|\Sigmav|^{1/2}}f_\mathcal{N}\left(\yb; \, \muv_{t+1}(\tb), \, \frac{1}{2}\Sigmav + \mathbf{S}_{t}(\tb) - \PHI_t(\tb, \ct)\right) f_\mathcal{N}\left(\muv_{t+1}(\tb); \, \muv_{t}(\tb), \, \PHI_t(\tb, \ct)\right) d\muv_{t+1}(\tb) \\& - \int \left(f_\mathcal{N}\left(\yb; \, \muv_{t+1}(\tb), \, \Sigmav + \mathbf{S}_{t}(\tb) - \PHI_t(\tb, \ct)\right)\right)^2 f_\mathcal{N}\left(\muv_{t+1}(\tb); \, \muv_{t}(\tb), \, \PHI_t(\tb, \ct)\right) d\muv_{t+1}(\tb).
        \end{split}
    \end{align}

The rest of the proof follows from \cite{Surer2023}, and we provide the remainder for completeness. For brevity, we omit $\tb$ from the notations $\muv_{t}(\tb)$ and $\muv_{t+1}(\tb)$. Defining $\mathbf{L} \coloneqq \frac{1}{2}\Sigmav + \Sv_t(\tb) - \PHI_t(\tb, \ct)$, $\mathbf{M} \coloneqq  \Sigmav + \Sv_t(\tb) - \PHI_t(\tb, \ct)$, and $a_1 \coloneqq \frac{2^{-d}\pi^{-d/2}|\Sigmav|^{-1/2}}{(2\pi)^d|\mathbf{L}\PHI_t(\tb, \ct)|^{1/2}}$, $a_2 \coloneqq \frac{(2\pi)^{-3d/2}}{|\mathbf{M}\PHI_t(\tb, \ct)\mathbf{M}|^{1/2}}$, and assuming $\mathbf{L}$ and $\mathbf{M}$ are invertible, \eqref{eq:IVAR_explore} is equivalently written as 
    \begin{align} 
        \begin{split} \label{eq:gnew}
        & a_1 \int \exp\left\{-\frac{1}{2} \left(\left(\yb - \muv_{t+1}\right)^\top \mathbf{L}^{-1} \left(\yb - \muv_{t+1}\right) + \left(\muv_{t+1} - \muv_{t}\right)^\top \PHI_t(\tb, \ct)^{-1} \left(\muv_{t+1} - \muv_{t}\right) \right)\right\}d\muv_{t+1} \\
        & - a_2 \int \exp\left\{-\frac{1}{2} \left(2\left(\yb - \muv_{t+1}\right)^\top \mathbf{M}^{-1} \left(\yb - \muv_{t+1}\right) + \left(\muv_{t+1} - \muv_{t}\right)^\top \PHI_t(\tb, \ct)^{-1} \left(\muv_{t+1} - \muv_{t}\right) \right)\right\}d\muv_{t+1}.
        \end{split}
    \end{align}
    Letting $\mathbf{v} \coloneqq \muv_t - \muv_{t+1}$ and $\mathbf{z} \coloneqq \yb - \muv_t$, and writing Equation~\eqref{eq:gnew} in matrix notation yields
    \begin{align}\label{jointlikematrix} 
            \begin{split}
                =& \frac{1}{2^{d}\pi^{d/2}|\Sigmav|^{1/2}} \int f_\mathcal{N}\left(\left[{\begin{array}{c} \mathbf{v} \\
                \mathbf{z} \\\end{array}} \right]; \,  \mathbf{0}, \, \left[ {\begin{array}{cc} \PHI_t(\tb, \ct) & -\PHI_t(\tb, \ct)\\
                -\PHI_t(\tb, \ct) & \mathbf{L} + \PHI_t(\tb, \ct)\\
                \end{array} } \right]  \right)d \mathbf{v} \\
                &-\frac{1}{2^{d}\pi^{d/2}|\mathbf{M}|^{1/2}} \int f_\mathcal{N}\left(\left[{\begin{array}{c} \mathbf{v} \\
                \mathbf{z} \\\end{array}} \right]; \,  \mathbf{0}, \, \left[ {\begin{array}{cc} \PHI_t(\tb, \ct) & -\PHI_t(\tb, \ct)\\
                -\PHI_t(\tb, \ct) & \frac{1}{2}\mathbf{M} + \PHI_t(\tb, \ct)\\
                \end{array} } \right]  \right)d \mathbf{v}.
            \end{split}
        \end{align}
    Marginalizing over $\mathbf{v}$ yields $\Exp\left(\VE\left[\tilde{p}(\yb|\tb)\right]\right)$ as
    \begin{align} \notag 
        \begin{split}
            \frac{f_\mathcal{N}\left(\yb; \, \muv_{t}(\tb), \, \frac{1}{2}\Sigmav + \Sv_t(\tb) \right)}{2^{d}\pi^{d/2}|\Sigmav|^{1/2}} - \frac{f_\mathcal{N}\left(\yb; \, \muv_{t}(\tb), \, \frac{1}{2}\left(\Sigmav + \Sv_t(\tb) + \PHI_t(\tb, \ct)\right)\right)}{2^{d}\pi^{d/2}|\Sigmav + \Sv_t(\tb) - \PHI_t(\tb, \ct)|^{1/2}} .
        \end{split}
    \end{align}

The IVAR criterion, ${\rm IVAR}(\ct, \breve{a})$, can be expressed as $\int\limits_{\tb \in \Theta} \Exp\left(\VE\left[\tilde{p}(\tb|\yb)\right]\right) d\tb$, which simplifies to
    \begin{gather} 
        \int\limits_{\tb \in \Theta} p(\tb)^2\left(\frac{f_\mathcal{N}\left(\yb; \, \muv_t(\tb), \, \frac{1}{2}\Sigmav + \Sv_t(\tb)\right)}{2^d \pi^{d/2} |\Sigmav|^{1/2}} -  \frac{f_\mathcal{N}\left(\yb; \, \muv_t(\tb), \, \frac{1}{2}\left(\Sigmav + \Sv_t(\tb) + \PHI_{t}(\tb, \ct)\right)\right)}{2^d \pi^{d/2} \left|\Sigmav + \Sv_t(\tb) - \PHI_{t}(\tb, \ct)\right|^{1/2}}\right)d\tb. \notag
    \end{gather}

%% file: appE.tex
\section{Details for Experiment Designs}
\label{sec:expdesign}

\begin{figure}[!h]
    \centering
    \begin{subfigure}{1\textwidth}    \centering
        \includegraphics[width=0.45\textwidth]{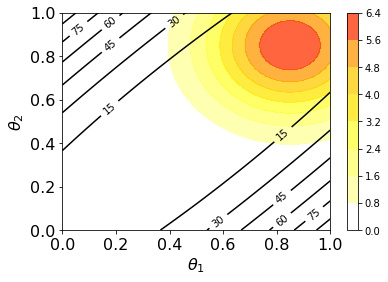}
    \end{subfigure}
    \begin{subfigure}{0.45\textwidth}
        \includegraphics[width=1\textwidth]{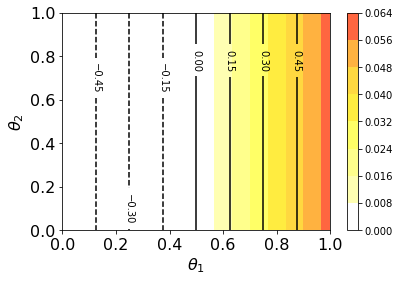}
    \end{subfigure}
    \begin{subfigure}{0.45\textwidth}
        \includegraphics[width=1\textwidth]{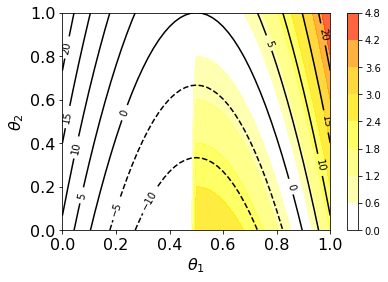}
    \end{subfigure}
    \begin{subfigure}{0.45\textwidth}
        \includegraphics[width=1\textwidth]{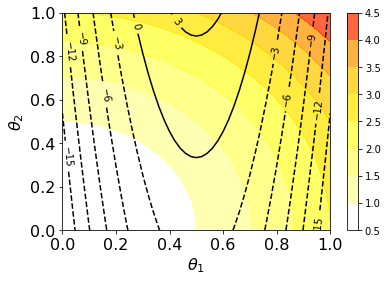}
    \end{subfigure}
    \begin{subfigure}{0.46\textwidth}
        \includegraphics[width=1\textwidth]{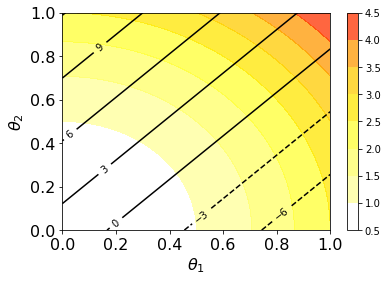}
     \end{subfigure}
    \caption{Illustrations of three synthetic simulation models: unimodal (first row), banana (second row), and bimodal (third row). Colors show the $j$the diagonal element of $\Rb(\cdot)$ and contour lines of the mean surface $\eta(\xb_j, \cdot)$ are overlaid for $j = 1, \ldots, d$.}
    \label{fig:synth_figs}
\end{figure}

This section describes the test functions employed in our numerical experiments. Figure~\ref{fig:synth_figs} shows the mean surface $\latent(\tg)$ and the intrinsic variance $\Rb(\tg)$ for each test function.

For the unimodal function with $p = 2$ and $d = 1$, we consider $\tg = (\vartheta_1, \vartheta_{2}) \in [0, 1]^{2}$, $\eta(\tg) = 0.26 \times \left(\left(20 \vartheta_1 - 10\right)^2 + (20 \vartheta_2 - 10)^2\right) - 0.48 \left(20 \vartheta_1 - 10\right) (20 \vartheta_2 - 10)$, and $R(\tg) = 2 \times f_\mathcal{N}(\tg; \mathbf{b}, \mathbf{C})$ with $\mathbf{b} = (0.85, 0.85)^\top$ and $\mathbf{C} = 0.05 \mathbf{I}_2$. The field data is generated through $y = \eta\left(\tb = \tb^*\right) + \epsilon$ with $\epsilon \sim {\rm N}(0, 0.1^2)$ and $\tb^* = (0.5, 0.5)^\top$.

For the banana function with $p = 2$ and $d = 2$, we consider $\tg = (\vartheta_1, \vartheta_{2}) \in [0, 1]^{2}$, $\latent(\tg) = \left(0.03\left(40 \vartheta_1 - 20\right), 15\vartheta_2 - 15 + 0.06 \left(40 \vartheta_1 - 20\right)^2\right)$. For the regions where $\vartheta_1 < 0.5$, $\Rb(\tg)$ is a $2 \times 2$ diagonal matrix with diagonal elements $0.01 |\latent(\tg)|$. In the regions where $\vartheta_1 \geq 0.5$, $\Rb(\tg)$ is also a $2 \times 2$ diagonal matrix, but with diagonal elements $0.1 \left|[\latent(\tg)]_{1}\right|$ and $0.2 \left|[\latent(\tg)]_{2}\right|$. The field data is generated through $\yb = \latent\left(\tb = \tb^*\right) + \epsilonv$ with $\epsilonv \sim {\rm MVN}(\mathbf{0}, \Sigmav)$ and $\tb^* = (0.5, 0.75)^\top$. $\Sigmav$ is a diagonal matrix with diagonal elements 0.03 and 0.5.

For the bimodal function with $p = 2$ and $d = 2$, we consider $\tg = (\vartheta_1, \vartheta_{2}) \in [0, 1]^{2}$, $\latent(\tg) = \left(\sqrt{0.2}\left(\left(12\vartheta_2-4\right) - \left(12\vartheta_1-6\right)^2\right), \sqrt{0.75}\left(\left(12\vartheta_2-4\right) - \left(12\vartheta_1-6\right)\right)\right)$, and $\Rb(\tg) = (0.5 + 2(\vartheta_1^2 + \vartheta_2^2))\mathbf{I}_2$. The field data is generated through $\yb = \latent\left(\tb = \tb^*\right) + \epsilonv$ with $\epsilonv \sim {\rm MVN}(\mathbf{0}, \Sigmav)$ and $\tb^* = (2/3, 2/3)^\top$. $\Sigmav$ is a diagonal matrix with diagonal elements 0.5.

%% file: appF.tex
\section{Assessing the Benefit of Replications}
\label{sec:experiment_replication}

\begin{figure}[h]
\centering
    \begin{subfigure}{1\textwidth}    
        \includegraphics[width=1\textwidth]{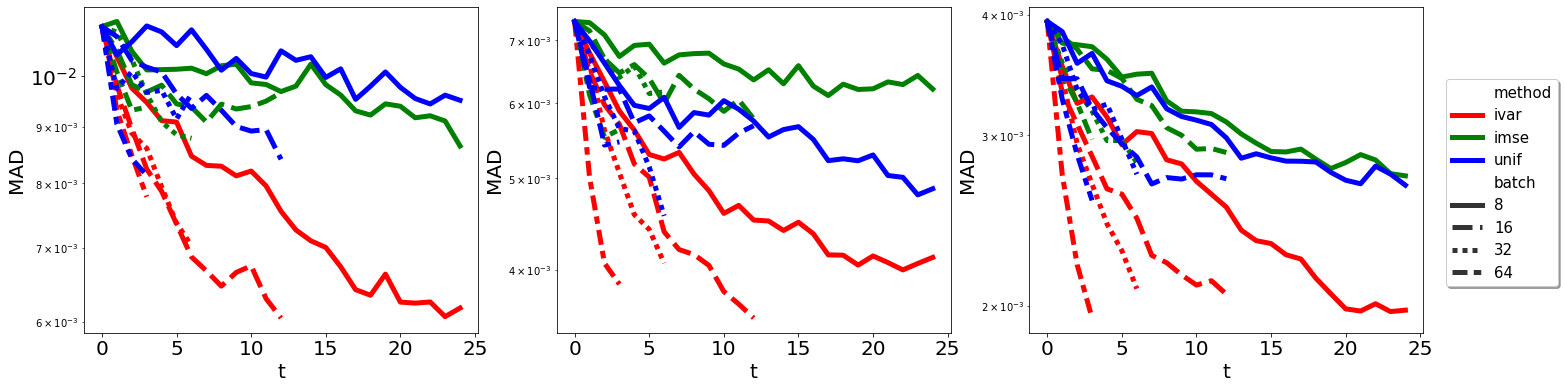}
    \end{subfigure}
    \begin{subfigure}{1\textwidth}    
        \includegraphics[width=1\textwidth]{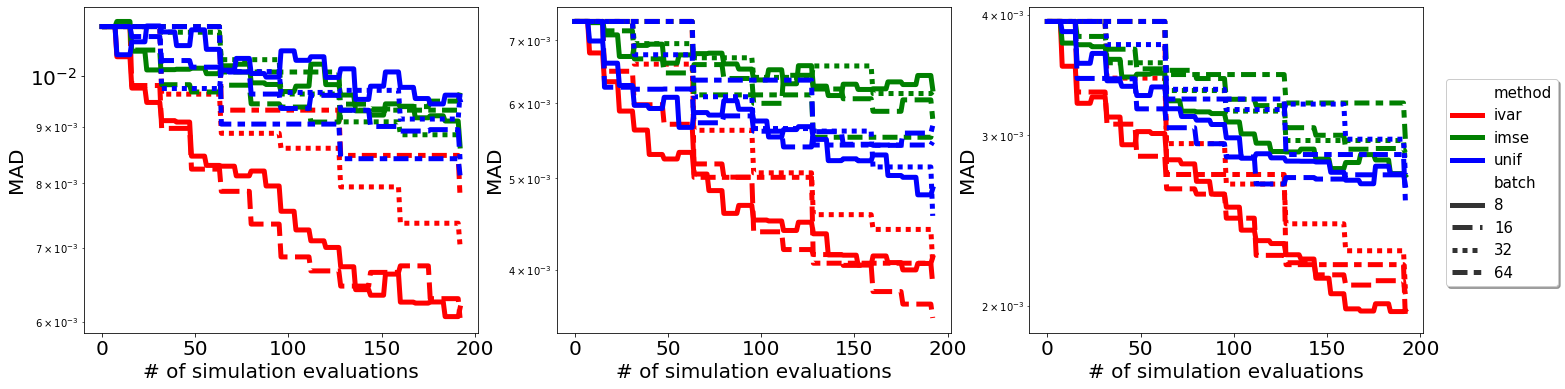}
    \end{subfigure}
    \caption{Comparison of different acquisition functions for replication with three synthetic simulation models: unimodal (left), banana (middle), and bimodal (right). Rows compare acquisition functions from two different perspectives.}
    \label{fig:results_rep}
\end{figure}
We analyze the benefits of replication around the region of interest using the proposed IVAR-based allocation strategy in \eqref{eq:ivar_allocation}, applied to unimodal, banana, and bimodal functions. 
We integrate the IMSE acquisition function in Supplementary Material~\ref{proof:imse_exploit} into our iterative framework for comparative purposes. We note that IMSE is used here to improve predictions of simulation outputs, rather than predictions of field data.
We assess the performance of batch-sequential approaches with batch sizes of $b \in \{8, 16, 32, 64\}$ to allocate a total of 192 ($=T_b \times b$) new simulation evaluations. 
For the initial sample, we create a $10 \times 10$ grid ($n_0 = 100$) and assign two replicates per point ($a_1 = \cdots = a_{n_0} = 2$). 
As a baseline benchmark procedure, we randomly select 96 points from the grid and allocate two additional replicates to each point. 
This approach is referred to as UNIF in the benchmark. 
The same initial sample is used for all methods--IVAR, IMSE, and UNIF. 
Likewise, the observed data is rerandomized for each replication and the same observed data is used across different methods within each replication. 
As a performance measure, we calculate the mean absolute difference (MAD) between the estimated posterior and the true posterior $\tilde{p}(\tb|\yb)$ at unseen calibration parameters. 
This involves generating a set of reference calibration parameters $\Theta_{\rm ref}$ and assessing the performance using the metric ${\rm MAD}_t = \frac{1}{|\Theta{\rm ref}|}\sum\limits_{\tb \in \Theta_{\rm ref}} |\tilde{p}(\tb|\yb) - \hat{p}_t(\tb|\yb)|$ at each stage $t$. 
Here, $\hat{p}_t(\tb|\y)$ represents the posterior prediction in \eqref{expectedpostfinal}. 
The reference set, $\Theta_{\rm ref}$, is a two-dimensional grid of size $50 \times 50$. 
Additionally, we calculate the interval score $S_\alpha(l, u; \theta^*)$ \citep{gneiting2007strictly} to evaluate both the width of the parameters selected with each method and the coverage of the best parameterization in each experimental replicate. 
The interval score is defined as  
\begin{gather}\label{eq:interval_score}
   S_\alpha(l, u; \theta^*) = (u-l) + \frac{2}{\alpha} (l - \theta^*) \mathbbm{1}\{{\theta^* < l}\} + \frac{2}{\alpha}(\theta^* - u)  \mathbbm{1}\{{\theta^* > u}\}.
\end{gather}
Here, $l$ and $u$ represent the quantiles of acquired parameters at level $\frac{\alpha}{2}$ and $(1-\frac{\alpha}{2})$, respectively, and $\mathbbm{1}$ is an indicator function. 
In our experiments, we substitute the maximum likelihood estimate of the parameter in place of $\theta^*$ and use $\alpha = 0.10$. 
\begin{table}[h]
\scriptsize
    \centering
    \setlength{\tabcolsep}{3pt}
    \begin{tabular}{cc|cccc|cccc|cccc}
                           & & \multicolumn{4}{c}{unimodal} & \multicolumn{4}{c}{banana} & \multicolumn{4}{c}{bimodal} \\ \hline
                           & & 8 & 16 & 32 & 64 & 8 & 16 & 32 & 64 & 8 & 16 & 32 & 64  \\ \hline
  \multirow{2}{*}{\rotatebox[origin=c]{90}{IVAR}} & $\theta_1$ & 0.4 (0.4) & 0.4 (0.2) & 0.6 (0.3) & 0.6 (0.1) & 0.3 (0.1) & 0.3 (0.1) & 0.4 (0.1) & 0.4 (0.1) & 0.4 (0.4) & 0.4 (0.1) & 0.4 (0.1) & 0.5 (0.1) \\
                                                  & $\theta_2$ & 0.4 (0.8) & 0.4 (0.7) & 0.6 (0.2) & 0.6 (0.1) & 0.8 (0.2) & 0.8 (0.1) & 0.8 (0.1) & 0.8 (0.1) & 0.5 (0.1) & 0.6 (0.1) & 0.6 (0.1) & 0.7 (0.1) \\
  \multirow{2}{*}{\rotatebox[origin=c]{90}{IMSE}} & $\theta_1$ & 0.7 (1.0) & 1.8 (1.1) & 0.6 (0.6) & 0.7 (0.4) & 0.9 (2.4) & 1.1 (2.4) & 0.9 (2.5) & 0.9 (2.2) & 1.0 (0.0) & 1.0 (0.0) & 1.0 (0.0) & 1.0 (0.0) \\
                                                  & $\theta_2$ & 0.7 (1.0) & 1.8 (1.2) & 0.6 (0.8) & 0.6 (0.7) & 1.0 (0.0) & 1.0 (0.0) & 1.0 (0.0) & 1.0 (0.0) & 1.0 (0.0) & 1.0 (0.0) & 1.0 (0.0) & 1.0 (0.0) \\ 
 \multirow{2}{*}{\rotatebox[origin=c]{90}{UNIF}}  & $\theta_1$ & 1.0 (0.0) & 1.0 (0.0) & 1.0 (0.0) & 1.0 (0.0) & 1.0 (0.0) & 1.0 (0.0) & 1.0 (0.0) & 1.0 (0.0) & 1.0 (0.0) & 1.0 (0.0) & 1.0 (0.0) & 1.0 (0.0) \\
                                                  & $\theta_2$ & 1.0 (0.0) & 1.0 (0.0) & 1.0 (0.0) & 1.0 (0.0) & 1.0 (0.0) & 1.0 (0.0) & 1.0 (0.0) & 1.0 (0.0) & 1.0 (0.0) & 1.0 (0.0) & 1.0 (0.0) & 1.0 (0.0) \\ \hline
    \end{tabular}
    \caption{Median interval scores for parameters selected using different acquisition functions for unimodal (left), banana (middle), and bimodal (right) functions. Values in parentheses indicate standard deviations across 30 replicates.}
    \label{tab:interval_score_replicate}
\end{table}

\begin{figure}[!h]
\centering
    \begin{subfigure}{0.42\textwidth}    
        \includegraphics[width=1\textwidth]{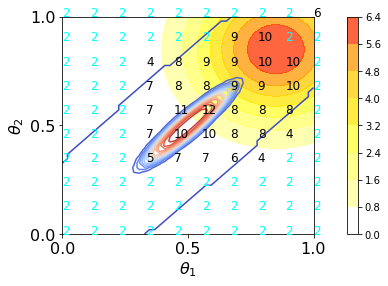}
    \end{subfigure}
    \begin{subfigure}{0.42\textwidth}    
        \includegraphics[width=1\textwidth]{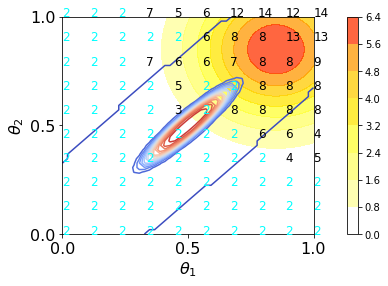}
    \end{subfigure}
    \begin{subfigure}{0.42\textwidth}
        \includegraphics[width=1\textwidth]{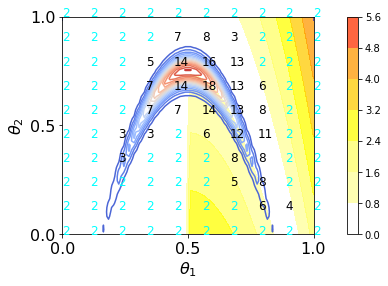}
    \end{subfigure}
    \begin{subfigure}{0.42\textwidth}
        \includegraphics[width=1\textwidth]{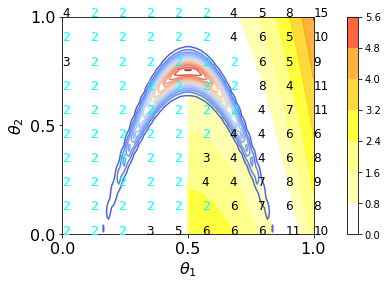}
    \end{subfigure}
    \begin{subfigure}{0.42\textwidth}
        \includegraphics[width=1\textwidth]{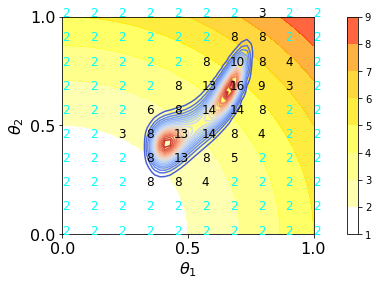}
    \end{subfigure}
    \begin{subfigure}{0.42\textwidth}
        \includegraphics[width=1\textwidth]{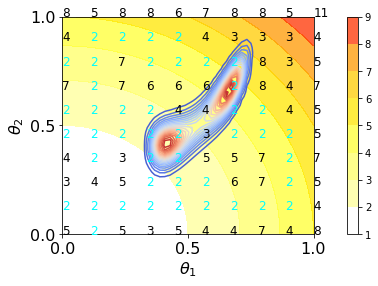}
     \end{subfigure}
    \caption{Allocation of 192 new replicates guided by IVAR (left) and IMSE (right) for unimodal (top row), banana (middle row), and bimodal (bottom row) functions. Contours show the posterior probability density. Numbers indicate the number of replicates at each parameter. Cyan markers represent the locations where no additional replicates are allocated. Background color indicates total intrinsic variance across $d$ dimensions.}
    \label{fig:synth_IVAR_illus}
\end{figure}    
Figure~\ref{fig:results_rep} summarizes MAD values averaged across 30 replicates. 
The first row displays MAD after each stage with the $x$-axis denoting stage $t$. 
The second row summarizes MAD following the acquisition of simulation runs with the $x$-axis representing the number of evaluations. 
The first row can approximate the total wall-clock time, especially if simulations constitute the primary computational overhead. 
Given that we acquire the same number of parameters with each batch size $b$, sequential methods with larger batch sizes terminate in fewer stages. 
Figure~\ref{fig:synth_IVAR_illus} demonstrates proposed allocations guided by IVAR and IMSE with $b = 32$ for a single replicate of the experiments. 
IVAR efficiently allocates replicates across both high and low posterior regions to improve learning of the posterior distribution. It effectively covers the parameter region of interest without wasting computational resources on simulations outside this region.
On the other hand, IMSE leads to inefficient use of computational resources by focusing on areas with high uncertainty outside the region of interest. 
For instance, IMSE allocates many replicates to the far right corner in the parameter space of the banana function, away from the region of interest, and does not allocate additional replicates to the left-hand side of the high posterior region due to low uncertainty. 
Therefore, if the calibration region of interest does not intersect with the high uncertainty region, the IMSE allocation offers little insight into the model's behavior within the parameter region of interest.

In all examples, IVAR achieves the smallest interval score, indicated in Table~\ref{tab:interval_score_replicate}, demonstrating its effectiveness in identifying the region of interest. 
Consequently, IVAR outperforms all competitors across various batch sizes.
The MAD metric shows no significant differences across different batch sizes. 
However, larger batches distribute replicates over a broader area, whereas smaller batches concentrate on more confined regions using both IVAR and IMSE. 
This trend is evident for IVAR in the increasing interval scores with larger batch sizes, as shown in Table~\ref{tab:interval_score_replicate}. 
Sequential procedures with larger batch sizes may terminate earlier, yet there is a risk of compromised accuracy from forming large batches without observing their simulation outputs. 
On the other hand, with smaller batch sizes, IVAR risks over-exploiting regions of high posterior probability with high uncertainty, while IMSE focuses exclusively on the regions of highest uncertainty.
For example, while IVAR achieves the lowest median interval score with a batch size of $b=8$ for the unimodal function, it exhibits the highest variability across 30 replicates compared to the remaining batches. 
This suggests inadequate coverage of the region of interest in some experimental replicates for the unimodal function. 
On the other hand, the MAD value increases when $b=64$ since fewer replicates are allocated to the calibration region of interest due to the expanded targeted area. 
Conversely, IMSE's MAD value decreases with $b=64$ for the unimodal function as the replicates overlap more with the high posterior region, reflecting IMSE's focus on a broader targeted area.
Consequently, practitioners need to select the optimal batch size depending on the specific application and the availability of the computational resources.

%% file: appG.tex
\section{Supporting Results for Synthetic Simulations}
\label{sec:additional_synt_res}
\begin{figure}[!h]
\centering
    \begin{subfigure}{1\textwidth}    
        \includegraphics[width=1\textwidth]{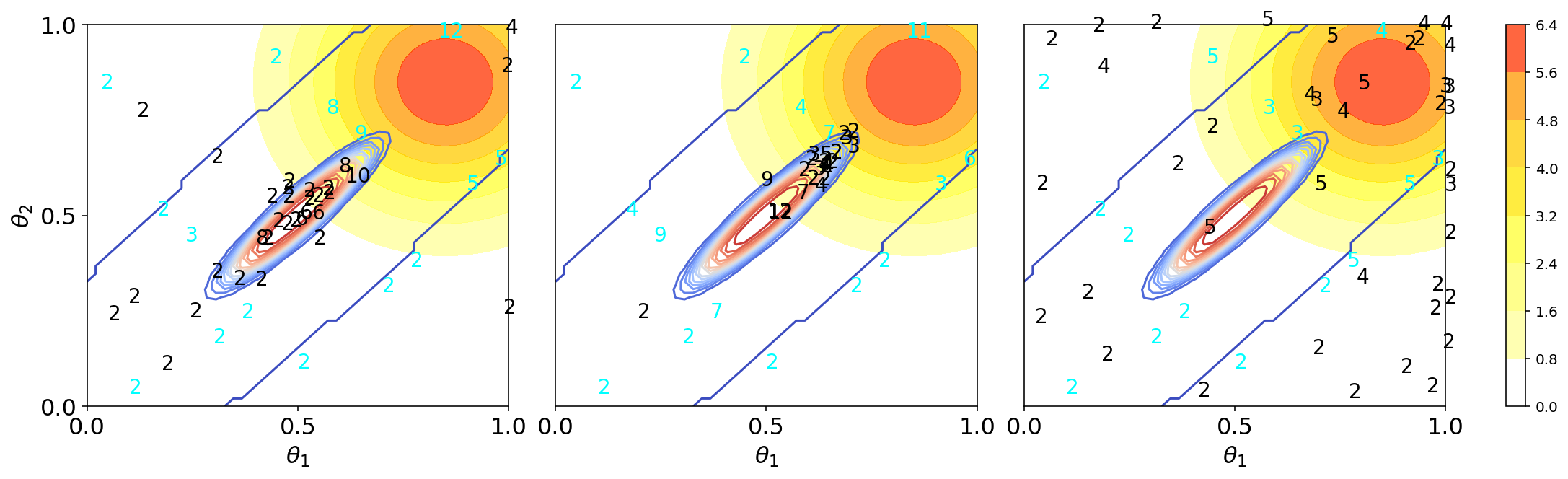}
    \end{subfigure}
    \begin{subfigure}{1\textwidth}    
        \includegraphics[width=1\textwidth]{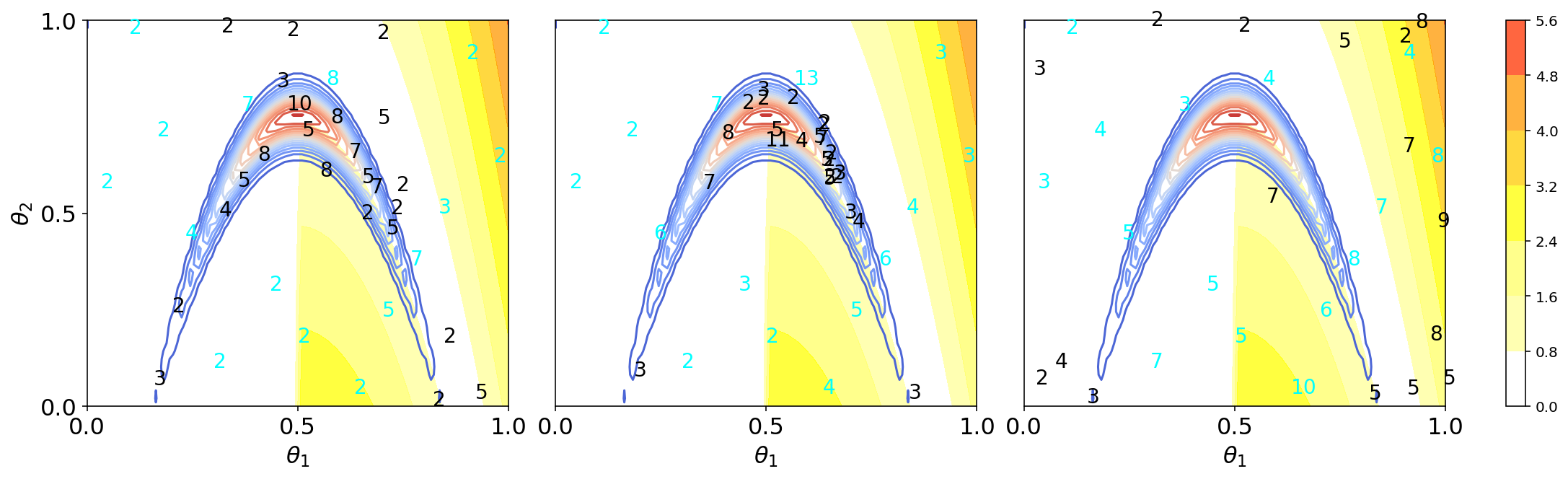}
    \end{subfigure}
    \begin{subfigure}{1\textwidth}
        \includegraphics[width=1\textwidth]{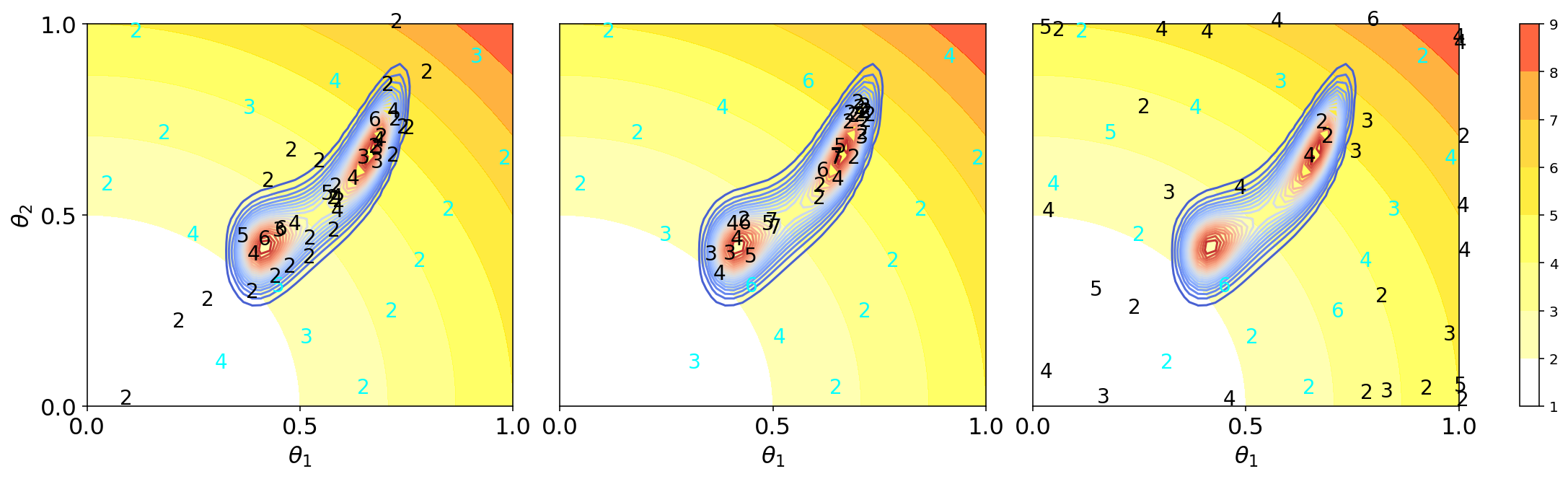}
    \end{subfigure}
    \caption{Allocation of 128 new simulation runs using IVAR (left), VAR (middle), and IMSE (right) for unimodal (top row), banana (middle row), and bimodal (bottom row) functions. Contours show the posterior probability density. Numbers indicate the number of replicates at each parameter. Cyan markers show initial sample locations ($n_0 = 15$). Background color indicates total intrinsic variance across $d$ dimensions.} 
    \label{fig:synth_IVAR_illus_hybrid}
\end{figure}

\begin{table}[ht]
\scriptsize
    \centering
    \setlength{\tabcolsep}{3pt}
    \begin{tabular}{cc|cccc|cccc|cccc}
                           & & \multicolumn{4}{c}{unimodal} & \multicolumn{4}{c}{banana} & \multicolumn{4}{c}{bimodal} \\ \hline
                           & & 8 & 16 & 32 & 64 & 8 & 16 & 32 & 64 & 8 & 16 & 32 & 64  \\ \hline
  \multirow{2}{*}{\rotatebox[origin=c]{90}{IVAR}} & $\theta_1$ & 0.6 (0.1) & 0.6 (0.1) & 0.6 (0.1) & 0.6 (0.1) & 0.5 (0.1) & 0.5 (0.1) & 0.5 (0.1) & 0.5 (0.1) & 0.5 (0.1) & 0.5 (0.1) & 0.4 (0.1) & 0.5 (0.1) \\
                                                  & $\theta_2$ & 0.6 (0.1) & 0.6 (0.1) & 0.6 (0.1) & 0.6 (0.1) & 0.8 (0.1) & 0.8 (0.1) & 0.8 (0.1) & 0.8 (0.1) & 0.6 (0.1) & 0.6 (0.1) & 0.6 (0.1) & 0.6 (0.1) \\
 \multirow{2}{*}{\rotatebox[origin=c]{90}{VAR}} & $\theta_1$ & 0.5 (0.1) & 0.5 (0.2) & 0.5 (0.2) & 0.5 (0.2) & 0.4 (0.1) & 0.5 (0.1) & 0.5 (0.1) & 0.5 (0.1) & 0.4 (0.1) & 0.4 (0.1) & 0.4 (0.1) & 0.4 (0.1) \\
                                                  & $\theta_2$ & 0.5 (0.5) & 0.5 (0.1) & 0.5 (0.2) & 0.5 (0.1) & 0.8 (0.1) & 0.7 (0.1) & 0.8 (0.1) & 0.8 (0.1) & 0.5 (0.1) & 0.5 (0.1) & 0.6 (0.1) & 0.6 (0.1) \\                     
  \multirow{2}{*}{\rotatebox[origin=c]{90}{IMSE}} & $\theta_1$ & 0.9 (0.0) & 0.9 (0.0) & 0.9 (0.0) & 0.9 (0.0) & 1.0 (0.0) & 1.0 (0.0) & 0.9 (0.0) & 0.9 (0.0) & 1.0 (0.0) & 1.0 (0.0) & 1.0 (0.0) & 1.0 (0.0) \\
                                                  & $\theta_2$ & 0.9 (0.0) & 0.9 (0.0) & 0.9 (0.0) & 0.9 (0.0) & 1.0 (0.0) & 1.0 (0.0) & 1.0 (0.0) & 1.0 (0.0) & 1.0 (0.0) & 1.0 (0.0) & 1.0 (0.0) & 1.0 (0.0) \\ 
 \multirow{2}{*}{\rotatebox[origin=c]{90}{UNIF}}  & $\theta_1$ & 0.9 (0.0) & 0.9 (0.0) & 0.9 (0.0) & 0.9 (0.0) & 0.9 (0.0) & 0.9 (0.0) & 0.9 (0.0) & 0.9 (0.0) & 0.9 (0.0) & 0.9 (0.0) & 0.9 (0.0) & 0.9 (0.0) \\
                                                  & $\theta_2$ & 0.9 (0.0) & 0.9 (0.0) & 0.9 (0.0) & 0.9 (0.0) & 0.9 (0.2) & 0.9 (0.2) & 0.9 (0.2) & 0.9 (0.2) & 0.9 (0.0) & 0.9 (0.0) & 0.9 (0.0) & 0.9 (0.0) \\ \hline    
    \end{tabular}
    \caption{Median interval scores for parameters selected using different acquisition functions for unimodal (left), banana (middle), and bimodal (right) functions. Values in parentheses indicate standard deviations across 30 replicates.}
    \label{tab:interval_score_explore}
\end{table}
\begin{figure}[!ht]
\centering
    \begin{subfigure}{1\textwidth}    
        \includegraphics[width=1\textwidth]{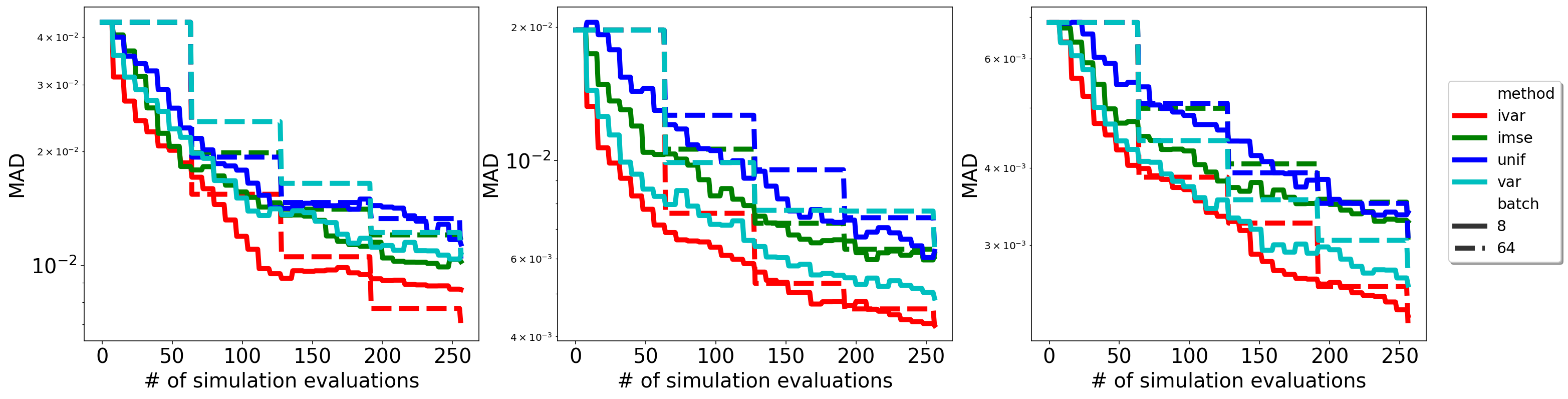}
    \end{subfigure}
    \caption{Comparison of different acquisition functions for unimodal (left), banana (middle), and bimodal (right) functions with $b \in \{8, 64\}$.}
    \label{fig:results_exps_2}
\end{figure}
To complement the experiments in Section~\ref{sec:synthetic}, this section provides additional results. Figure~\ref{fig:results_exps_2} offers an alternative perspective on the results shown in Figure~\ref{fig:results_exps}. Figure~\ref{fig:synth_IVAR_illus_hybrid} displays the acquired parameters from a single replicate of the IVAR, VAR, and IMSE experiments. Table~\ref{tab:interval_score_explore} summarizes the corresponding interval scores across 30 replicates.

%% file: appH.tex
\section{Computational Time Comparison}
\label{sec:computational_performance}

We evaluate the computational performance of IVAR across batch sizes $b \in \{8, 16, 32,$ $ 64\}$. These experiments use the bimodal function as an example and involve acquiring 320 parameters separately through replication and exploration strategies. For the replication strategy, the initial sample consists of a $10 \times 10$ grid, with two replicates assigned to each point. For the exploration strategy, we begin with an initial sample of 15 points generated via LHS, with each point replicated twice. To assess computational performance, we measure the duration required to acquire a batch of $b$ parameters at each stage $t$. 

\begin{figure}[!h]
\centering
    \begin{subfigure}{0.45\textwidth}    
        \includegraphics[width=1\textwidth]{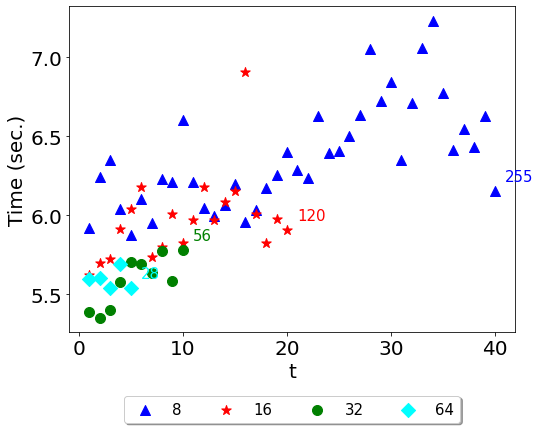}
    \end{subfigure}
    \begin{subfigure}{0.45\textwidth}    
        \includegraphics[width=1\textwidth]{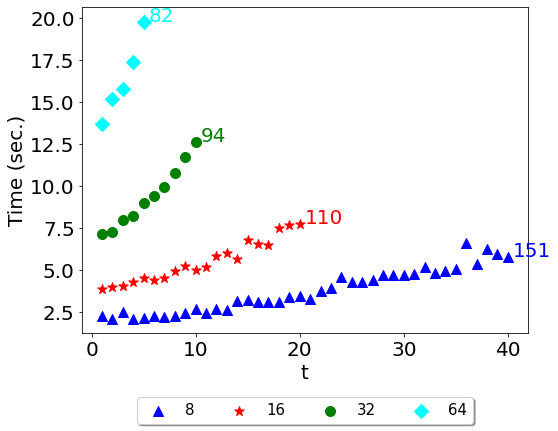}
    \end{subfigure}
    \caption{Computation times for the replication (left) and exploration (right) strategies. Annotated numbers indicate the total acquisition time for each batch size.}
    \label{fig:time_compare}
\end{figure}
Figure~\ref{fig:time_compare} presents the average acquisition times (in seconds) at each stage for batch size $b$, computed across ten replicates. In the replication strategy, the criterion \eqref{eq:ivar_allocation} integrates over the parameter space once for each unique parameter included in the design. Additionally, for both replication and exploration strategies, at each stage, the cost of building an independent emulator depends on the number of unique parameters (i.e., the computational expense of $\mathcal{O}(n_t^3)$). Since the set of unique parameters remains fixed throughout the procedure (i.e., $10 \times 10$ grid), the computational time for the replication strategy does not vary significantly across stages. It exhibits approximately linear scaling, with the total acquisition time decreasing proportionally as the batch size increases. In contrast, the exploration strategy involves iterating through the for loop in Algorithm~\ref{alg:krigingbeliever} $\breve{b}$ times. For each iteration, the criterion \eqref{eq:IVARexplore} requires integration over the parameter space for every candidate parameter in $\mathcal{L}_{\breve{t}}$. Moreover, the number of unique parameters increases at each stage. Consequently, the time required to acquire a batch of $b$ parameters increases at a faster rate with the exploration strategy compared to the replication strategy. Both \eqref{eq:ivar_allocation} and \eqref{eq:IVARexplore} involve integrals over a multi-dimensional parameter space, approximated by a sum over discrete sets $\Theta_{\rm ref}$. The size of $\Theta_{\rm ref}$ impacts the computational performance of both strategies. For higher-dimensional parameter spaces, methods such as sparse grids, smart sampling techniques, or advanced quadrature schemes can be employed to approximate the integrals more efficiently. Additionally, while the use of replication provides significant computational gains for each hetGP--reducing the computational complexity from $\mathcal{O}(N_t^3)$ to $\mathcal{O}(n_t^3)$, where $N_t = \sum_{i=1}^{n_t} a_i$--efficient emulators can also serve as replacements for the existing approach. Recent studies have introduced promising, highly efficient alternatives to traditional GPs, ranging from effective approximations \citep{wenger2024} to parallelizable exact emulators \citep{NoackExact2023}. Although these techniques are primarily designed for deterministic models, adapting similar approaches for stochastic models could provide more computationally effective replacements for conventional emulators. We also note that the computational expense of the acquisition function becomes comparatively less significant when the cost of obtaining a single simulation output is high, as the overall cost is primarily driven by the simulation runs rather than the acquisition step.

%% file: appI.tex
\section{Additional Results for Epidemiological Models}
\label{sec:epi_illustrate}

\begin{figure}[!h]
    \centering
    \begin{subfigure}{0.32\textwidth}  
        \includegraphics[width=1\textwidth]{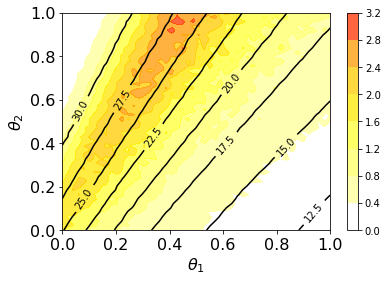}
    \end{subfigure}
    \begin{subfigure}{0.32\textwidth}
        \includegraphics[width=1\textwidth]{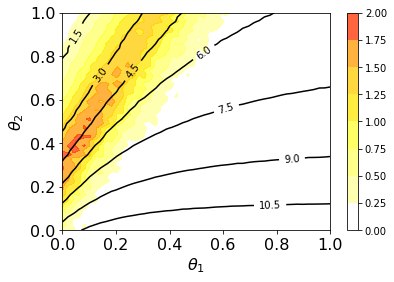}
    \end{subfigure}
    \begin{subfigure}{0.32\textwidth}
        \includegraphics[width=1\textwidth]{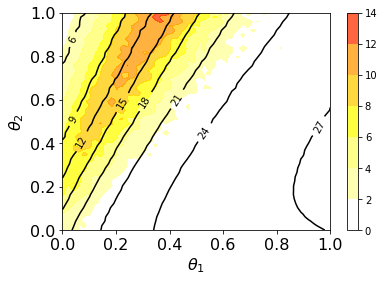}
    \end{subfigure}
    \caption{Illustration of the SIR model. Colors show the intrinsic variance and contour lines represent the mean surface of each simulation output. Each panel represents a dimension of the simulation output for $d = 3$.}
    \label{fig:synth_figs_SIR}
\end{figure}

This section provides further analysis with epidemiological models. Table~\ref{tab:sir_params} presents the parameters used in the SIR and SEIRDS models. Figure~\ref{fig:synth_figs_SIR} shows the mean surface $\latent(\tg)$ and the intrinsic variance $\Rb(\tg)$ for the SIR model. 
\begin{table}
    \footnotesize
    \centering
    \setlength{\tabcolsep}{3pt}
    \begin{tabular}{ccccc}
        & Parameter  & Label & Definition  & Prior \\ \hline
    \multirow{2}{*}{SIR} & $\beta$  & $\theta_1$ & infection rate & $\left[0.1, 0.3\right]$ \\
                         & $\gamma$  & $\theta_2$ & removal rate & $\left[0.05, 0.15\right]$ \\ \hline
    \multirow{7}{*}{SEIRDS} & $\beta$  & $\theta_1$ & rate of infection & $\left[0.15, 0.45\right]$ \\
                            & $\delta$  & $\theta_2$ & rate at which symptoms appear & $\left[0.15, 0.45\right]$ \\ 
                            & $\gamma_R$  & $\theta_3$ & recovery rate & $\left[0.04, 0.12\right]$ \\ 
                            & $\gamma_D$  & $\theta_4$ & death rate & $\left[0.06, 0.18\right]$ \\ 
                            & $\mu$  & $\theta_5$ & case fatality ratio & $\left[0.35, 1.00\right]$ \\ 
                            & $\epsilon$  & $\theta_6$ & import rate of infected individuals & $\left[0.05, 0.15\right]$ \\ 
                            & $\omega$  & $\theta_7$ & rate waning immunity & $\left[0.005, 0.015\right]$ \\ \hline
    \end{tabular}
    \caption{Epidemiological parameters and their prior ranges for SIR and SEIRDS models.}
    \label{tab:sir_params}
\end{table}

\begin{figure}[t]
\centering
    \begin{subfigure}{0.8\textwidth}    
        \includegraphics[width=1\textwidth]{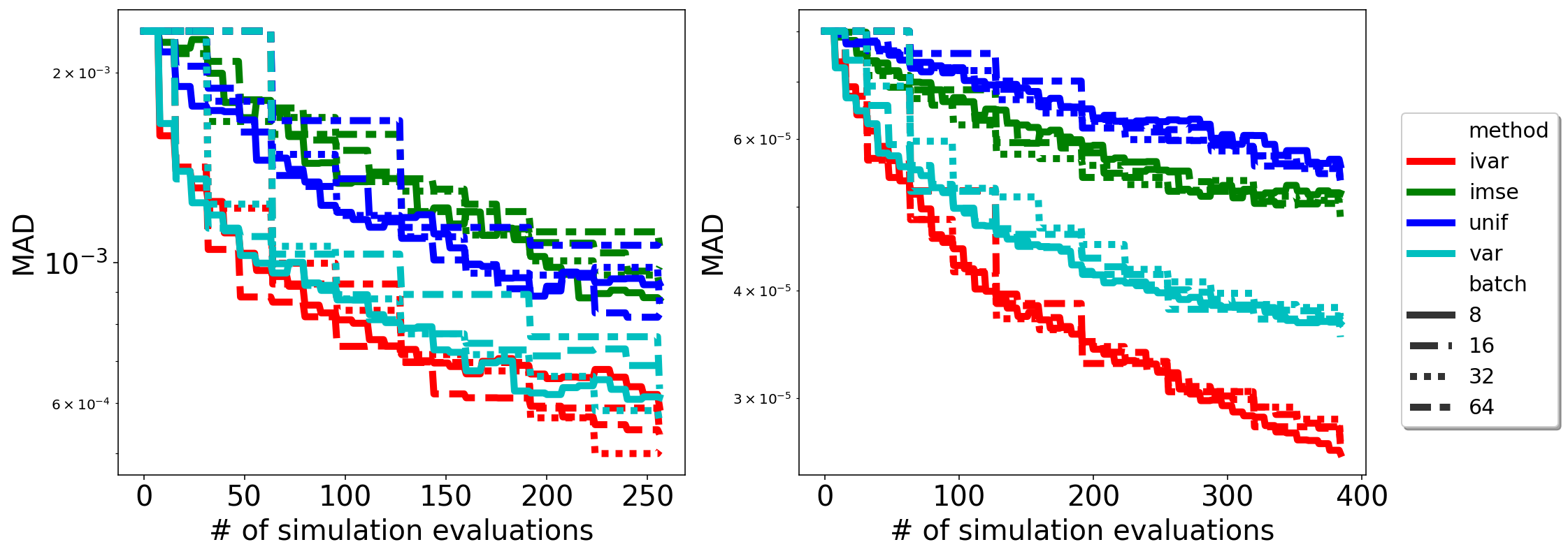}
    \end{subfigure}
    \caption{Comparison of different acquisition functions for SIR (left) and SEIRDS (right) models.}
    \label{fig:results_SIR_2}
\end{figure}
\begin{figure}[!h]
\centering
    \begin{subfigure}{1\textwidth}    
        \includegraphics[width=1\textwidth]{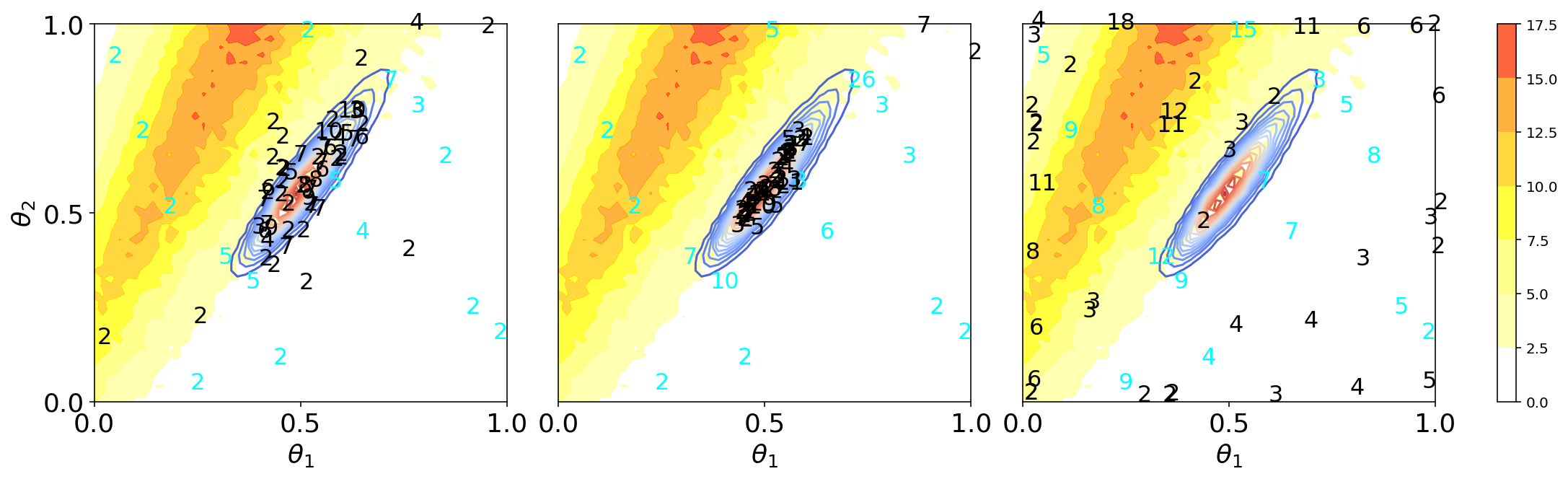}
    \end{subfigure}
    \caption{Illustration of acquired parameters using IVAR (left), VAR (middle), and IMSE (right). Contours show the posterior probability density. Colors show the total intrinsic variance across S, I, and R compartments. Numbers indicate the number of replicates at each parameter. Cyan markers show initial sample locations.}
    \label{fig:params_SIR}
\end{figure}
\begin{figure}[h]
\centering
    \begin{subfigure}{0.85\textwidth}    
        \includegraphics[width=1\textwidth]{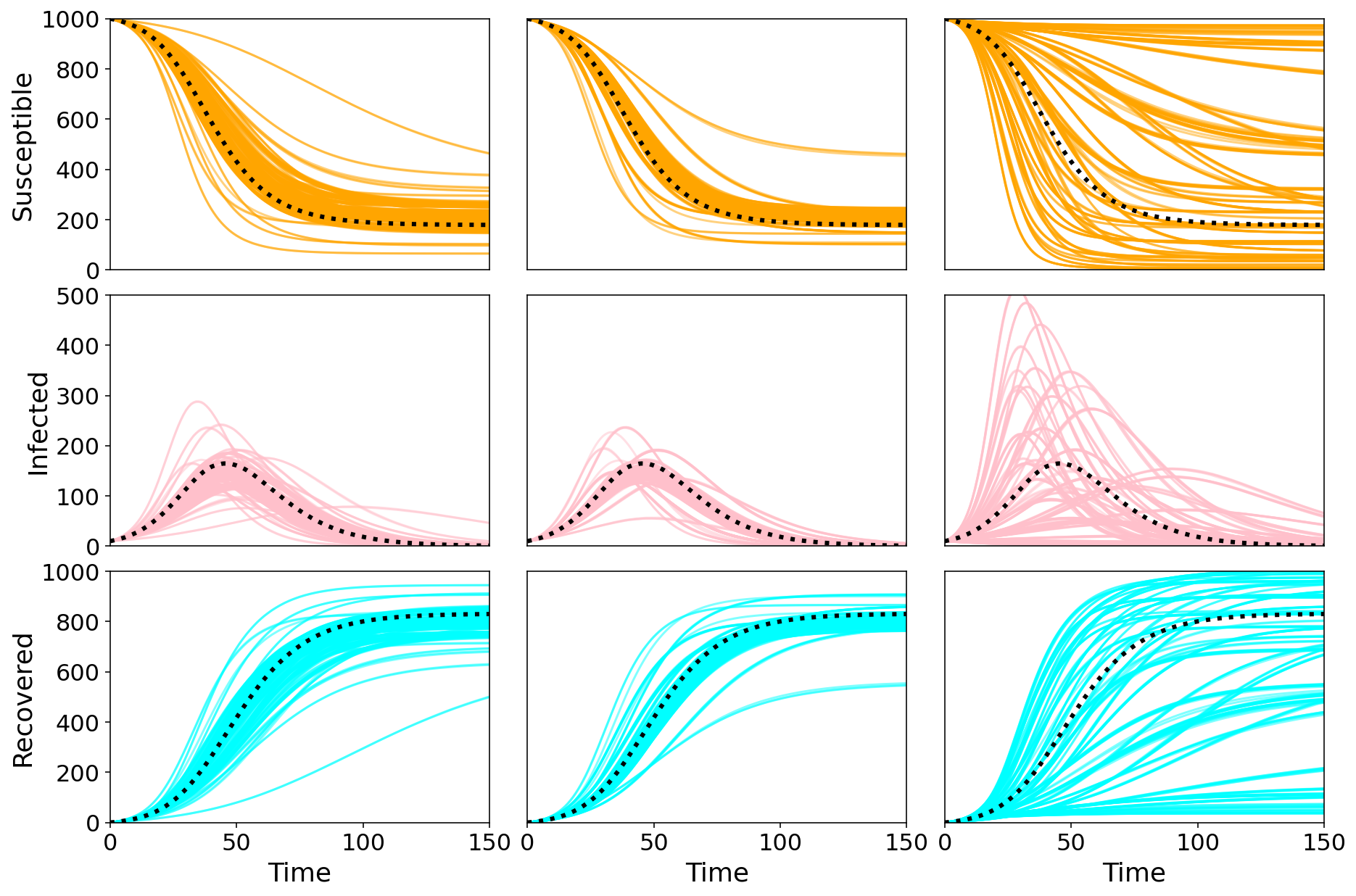}
    \end{subfigure}
    \caption{Simulation outputs of the SIR model obtained with the parameters acquired by IVAR (left), VAR (middle), and IMSE (right). The black dotted line denotes the expected simulation output at $\tb^*$.}
    \label{fig:outputs_SIR}
\end{figure}
\begin{figure}[t]
    \centering
    \begin{subfigure}{0.85\textwidth}  
        \includegraphics[width=1\textwidth]{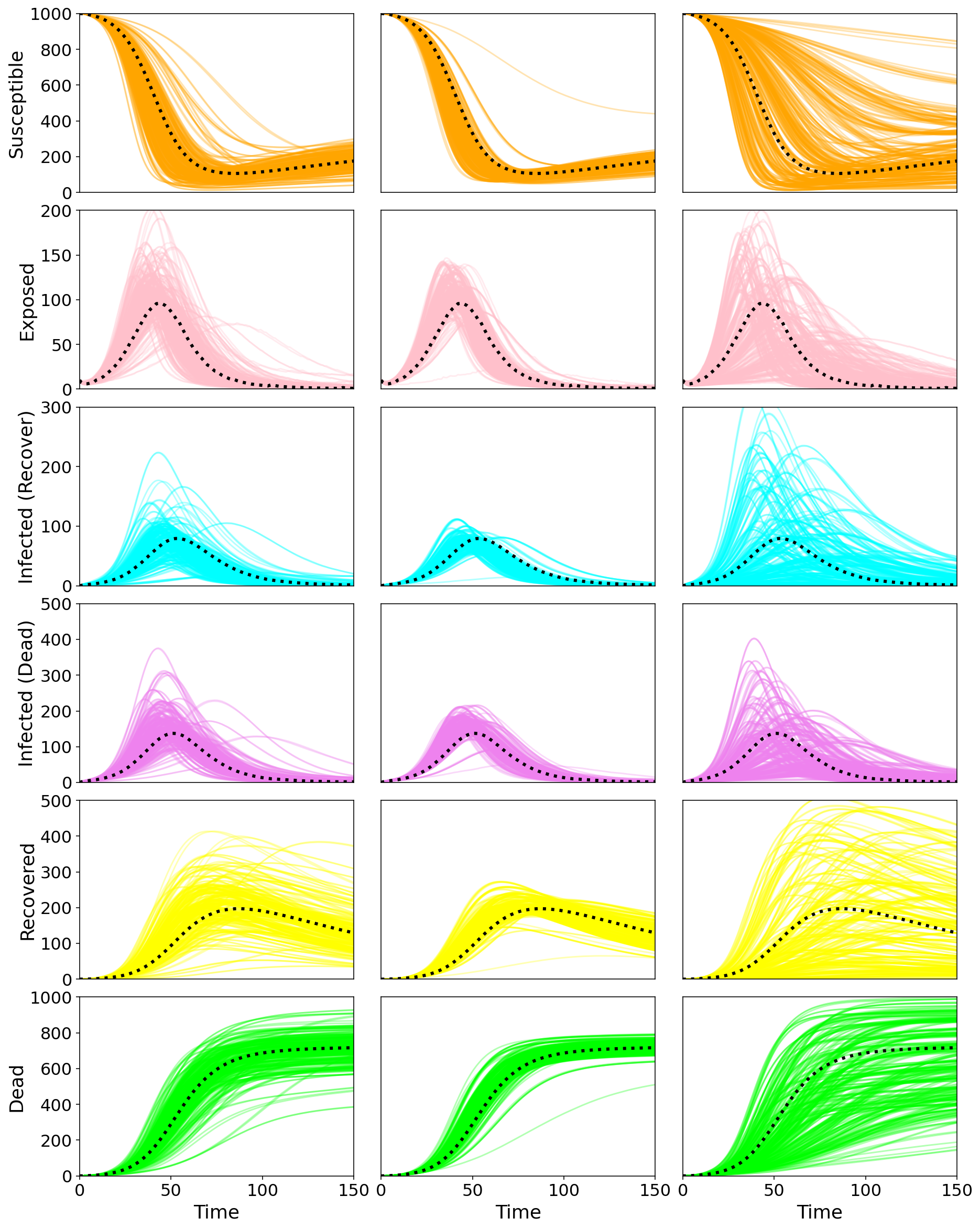}
    \end{subfigure}
    \caption{Simulation outputs of the SEIRDS model obtained with the parameters acquired by IVAR (left), VAR (middle), and IMSE (right). The black dotted line denotes the expected simulation output at $\tb^*$.}
    \label{fig:outputs_SEIRDS}
\end{figure}

Figure~\ref{fig:results_SIR_2} presents MAD values across 30 experimental replicates, offering an alternative perspective to Figure~\ref{fig:results_SIR}. Table~\ref{tab:interval_score_SIR} reports the interval scores of parameters selected using various acquisition functions for the SIR and SEIRDS models over 30 replicates. Figure~\ref{fig:params_SIR} shows the parameters obtained by IVAR, VAR, and IMSE in a single replicate of the SIR experiments. Figures~\ref{fig:outputs_SIR}--\ref{fig:outputs_SEIRDS} display the average simulation outputs across 1000 runs of the SIR and SEIRDS models, using parameters from a single replicate of each acquisition strategy.

\begin{table}[!ht]
\scriptsize
    \centering
    \setlength{\tabcolsep}{3pt}
    \begin{tabular}{cc|cccc|cccc|cccc}
                           & & \multicolumn{4}{c}{IVAR} & \multicolumn{4}{c}{VAR} & \multicolumn{4}{c}{IMSE}  \\ \hline
                           & & 8 & 16 & 32 & 64 & 8 & 16 & 32 & 64 & 8 & 16 & 32 & 64  \\ \hline
  \multirow{2}{*}{\rotatebox[origin=c]{90}{SIR}} & $\theta_1$  & 0.4 (0.0) & 0.4 (0.0) & 0.3 (0.0) & 0.4 (0.1) & 0.3 (0.1) & 0.3 (0.1) & 0.3 (0.1) & 0.3 (0.1) & 1.0 (0.0) & 1.0 (0.0) & 1.0 (0.0) & 1.0 (0.0) \\
                                                  & $\theta_2$ & 0.4 (0.1) & 0.5 (0.1) & 0.5 (0.1) & 0.5 (0.1) & 0.4 (0.1) & 0.4 (0.1) & 0.5 (0.1) & 0.4 (0.1) & 1.0 (0.0) & 1.0 (0.0) & 1.0 (0.0) & 1.0 (0.0)  \\
  \multirow{7}{*}{\rotatebox[origin=c]{90}{SEIRDS}} & $\theta_1$ & 0.5 (0.1) & 0.5 (0.1) & 0.5 (0.1) & 0.5 (0.1) & 0.4 (0.2) & 0.4 (0.1) & 0.4 (0.1) & 0.4 (0.1) & 0.9 (0.2) & 0.9 (0.1) & 0.9 (0.1) & 0.9 (0.1)  \\
                                                  & $\theta_2$   & 0.7 (0.1) & 0.7 (0.1) & 0.7 (0.1) & 0.7 (0.1) & 0.5 (0.1) & 0.5 (0.1) & 0.6 (0.1) & 0.5 (0.1) & 0.9 (0.0) & 0.9 (0.0) & 0.9 (0.0) & 0.9 (0.0)  \\ 
                                                  & $\theta_3$   & 0.8 (0.1) & 0.8 (0.1) & 0.8 (0.1) & 0.8 (0.1) & 0.7 (0.1) & 0.7 (0.1) & 0.7 (0.1) & 0.7 (0.1) & 0.9 (0.0) & 0.9 (0.2) & 0.9 (0.1) & 0.9 (0.1)  \\ 
                                                  & $\theta_4$   & 0.6 (0.1) & 0.7 (0.1) & 0.6 (0.1) & 0.6 (0.1) & 0.5 (0.1) & 0.5 (0.1) & 0.5 (0.1) & 0.5 (0.1) & 0.9 (0.0) & 0.9 (0.0) & 0.9 (0.0) & 0.9 (0.1)  \\ 
                                                  & $\theta_5$   & 0.4 (0.1) & 0.4 (0.1) & 0.5 (0.1) & 0.5 (0.1) & 0.2 (0.0) & 0.2 (0.0) & 0.2 (0.1) & 0.2 (0.0) & 0.9 (0.0) & 0.9 (0.0) & 0.9 (0.0) & 0.9 (0.0)  \\ 
                                                  & $\theta_6$   & 0.9 (0.0) & 0.9 (0.0) & 0.9 (0.0) & 0.9 (0.0) & 0.9 (0.1) & 0.9 (0.1) & 0.9 (0.0) & 0.9 (0.1) & 0.9 (0.2) & 0.9 (0.1) & 0.9 (0.1) & 0.9 (0.1)  \\ 
                                                  & $\theta_7$   & 0.9 (0.0) & 0.9 (0.0) & 0.9 (0.0) & 0.9 (0.0) & 0.9 (0.0) & 0.9 (0.0) & 0.9 (0.1) & 0.9 (0.1) & 0.9 (0.1) & 0.9 (0.1) & 0.9 (0.2) & 0.9 (0.1)  \\ \hline                                            
    \end{tabular}
    \caption{Median interval scores for parameters selected using different acquisition functions for SIR and SEIRDS models. Values in parentheses indicate standard deviations across 30 replicates.}
    \label{tab:interval_score_SIR}
\end{table}

%% file: appJ.tex
\clearpage

\section{Emulator Predictive Performance}
\label{sec:emuperformance}

\begin{table}[ht]
    \centering
    \footnotesize
    \setlength{\tabcolsep}{4pt} 
    \caption{Emulator performance for the SIR and SEIRDS models via MAPE values.}
    \label{tab:example}
    \begin{tabular}{c|cccccc|cccccc}
        \hline
                             &  \multicolumn{6}{|c|}{Emulator mean} &  \multicolumn{6}{|c}{Intrinsic variance} \\ \hline
        \diagbox{Example}{$j$} & 1 & 2 & 3 & 4 & 5 & 6 & 1 & 2 & 3 & 4 & 5 & 6\\ \hline
        SIR                  & 0.0 & 1.0 & 1.0 & N/A & N/A  & N/A & 13.0 & 25.0 & 17.0 & N/A & N/A & N/A \\ 
        SEIRDS               & 1.0 & 1.0 & 8.0 & 2.0 & 7.0 & 1.0 & 14.0 & 21.0 & 16.0 & 17.0 & 15.0 & 17.0       \\ \hline
    \end{tabular}
    \label{tab:emu_quality}
\end{table}
This section evaluates the emulator's accuracy for high-dimensional outputs using real-data experiments from the SIR and SEIRDS models. The SIR model produces three-dimensional outputs, while the SEIRDS model generates six-dimensional outputs. For the SIR model, we use a $50 \times 50$ grid of two-dimensional parameters as the reference data set, $\Theta_{\rm ref}$. For the SEIRDS model, the reference set consists of 2500 parameters sampled via LHS. Each parameter in the reference set is used to replicate the simulation model 1000 times. The average of these replicates serves as the expected simulation output, while the variance across replicates represents the true intrinsic variance. As training data, we generate 200 unique parameters for each of the SIR and SEIRDS models using LHS, with each parameter replicated 50 times. To assess emulator accuracy, we generate 30 training data sets. For each data set, the emulator is trained and then used to predict outputs for the reference set. For each output $j$, we measure the accuracy of the emulator's mean prediction using the mean absolute percentage error (MAPE), defined as ${\rm MAPE} = 100\frac{1}{|\Theta_{\rm ref}|}\sum\limits_{\tg \in \Theta_{\rm ref}}\left|\frac{\eta(\xb_j, \tg) - m_{t,j}(\tg)}{\eta(\xb_j, \tg)}\right|$, where $\eta(\xb_j, \tg)$ is the expected simulation output and $m_{t,j}(\tg)$ is the emulator prediction. Similarly, to evaluate the quality of intrinsic variance estimates, we compute MAPE by replacing $\eta(\xb_j, \tg)$ and $m_{t,j}(\tg)$ with the true and estimated intrinsic variance values. Table~\ref{tab:emu_quality} presents the median MAPE values across 30 experiment replicates. Emulators for mean prediction achieve consistently very low MAPE values  (i.e., $< 10 \%$ in all cases), demonstrating very high accuracy for both the SIR and SEIRDS models. While MAPE values are higher when estimating intrinsic noise, the emulator still performs well (i.e., $\leq 25 \%$ in all cases) in capturing it for both models.